# Molecular Junctions for Terahertz Switches and Detectors.

Imen Hnid,[1*] Ali Yassin,[2,3] Imane Arbouch,[4] David Guérin,[1] Colin van Dyck,[5]

Lionel Sanguinet,[2] Stéphane Lenfant,[1] Jérôme Cornil,[4]

Philippe Blanchard[2] and Dominique Vuillaume.[1*]

*1. Institute for Electronics Microelectronics and Nanotechnology (IEMN), CNRS, University of Lille, Av. Poincaré, Villeneuve d'Ascq, France.*
*2. MOLTECH-Anjou, CNRS, University of Angers, SFR MATRIX, F-49000 Angers, France.*
*3. Natural Sciences Department, School of Arts and Sciences, Lebanese American University, Beirut, Lebanon.*
*4. Laboratory for Chemistry of Novel Materials, University of Mons, Belgium.*
*5. Theoretical Chemical Physics group, University of Mons, Mons, Belgium.*

**Abstract:** Molecular electronics targets tiny devices exploiting the electronic properties of the molecular orbitals, which can be tailored and controlled by the chemical structure/conformation of the molecules. Many functional devices have been experimentally demonstrated; however, these devices were operated in the low frequency domain (mainly, dc to MHz). This represents a serious limitation for electronic applications, albeit molecular devices working in the THz regime have been theoretically predicted. Here, we experimentally demonstrate molecular THz switches at room temperature. The devices consist of self-assembled monolayers of molecules bearing two conjugated moieties coupled through a non-conjugated linker. These devices exhibit clear negative differential conductance behaviors (peaks in the current-voltage curves), as confirmed by *ab initio* simulations, which were reversibly suppressed under illumination with a 30 THz wave. We analyze how the THz switching behavior depends on the THz wave properties (power, frequency), and we benchmark that these molecular devices would outperform actual THz detectors.

At the nanoscale, the interactions between molecular-based devices and microwave-to-terahertz electromagnetic waves are a fascinating and relatively recent field of research since the beginning of molecular-scale electronics fifty years ago.[1, 2] As examples, the polarizability of alkyl chains and π-conjugated oligomers in self-assembled monolayers (SAMs) were first measured by combining a dc bias and a microwave signal (around few to few tens of GHz) in a rf-STM (radio frequency scanning tunneling microscope) experiment.[3, 4] The spin resonance of single molecules deposited on surfaces can be probed and manipulated (in UHV and at low temperatures) with a rf-STM, combined with a magnetic field (ESR-STM: electron spin resonance STM).[5-8] THz waves coupled with a STM (lightwave-driven THz-STM)[9, 10] was proven a powerful approach to study ultrafast dynamics of single molecules (*e.g.* ultrafast electron transfer), from simple $H_2$ to pentacene, nickelocene, for example.[11-13] These works have demonstrated their benefit to overcome the diffraction limits allowing to reach simultaneously high spatial and ultrafast temporal resolution affording new insights for nanoscale devices (see a review in Ref. 9). However, to the best of our knowledge, none of these works concern the direct measurement of the dynamic conductance of molecular devices in this microwave frequency regime and at room temperature.

From a device point of view, molecular-scale electronics suffers a serious limitation for electronic applications, because their electronic transport properties have been mainly studied in the dc regime, or restricted to low-frequency (< MHz). However, most of the metrics of these molecular devices (*e.g.,* injection barrier height at the electrode interfaces, electron transit time…) scale in the THz frequency range. For example, transit times through molecular junctions of a few femtoseconds (*i.e.,* THz frequency range) were theoretically predicted.[14, 15] Trasobares *et al.* demonstrated that a molecular diode can be operated at a frequency of ~18 GHz, and a frequency bandwidth of 520 GHz was estimated from a simple device modeling.[16] The electron transport properties of single molecule ($C_{60}$) transistors can also be tuned by THz radiation. THz photon-assisted tunneling and electron-vibron coupling were also recently demonstrated.[17, 18] Tunneling charge transfer plasmon was demonstrated at 145 and 244 THz, depending on the nature of the molecules in the molecular junction.[19] In this context, a molecular terahertz switch was theoretically proposed but never experimentally demonstrated. The suggested device is based on the dynamic interaction between a THz wave and a molecular junction featuring a negative differential conductance (NDC).[20]

Here, we demonstrate such a molecular THz switch device. We study the electron transport properties under 2.5-30 THz irradiation of SAMs of molecules (dihydroanthracene derivatives) made of two conjugated parts coupled through a non-conjugated linker (generally referenced as π-σ-π system). Due to this non-conjugated linker, the molecule can be seen as two weakly coupled sites in series constituting an efficient conducting channel only when the energies of the two sites are aligned leading to NDC behaviors.[21-24] We demonstrate that the NDC behavior is reversibly suppressed (observed) when the THz irradiation is turned on (off). We establish how this switching behavior depends on the THz wave characteristics (frequency, power). To explain these features, mechanisms involving THz-induced resonant electron transfer between the neighboring subunits, "coherent destruction of tunneling" in nanodevices[25] and molecular junctions,[26-28] or interaction between a surface plasmon polariton (SPP) and two-level molecular devices,[29, 30] are considered. Finally, from a benchmarking of the device performances, we envision that these THz molecular switches are prone for applications as THz detectors.

The molecules are based on a dihydroanthracene core which was symmetrically linked to two π-conjugated ethynylthiophene units. Both compounds were end-capped with two thiol functions protected by 2-cyanoethyl or 2-(trimethylsilyl)ethyl groups referred to as AH1 and AH2 (Fig. 1a), respectively (details in the Supporting Information). A third molecule already known to exhibit a NDC behavior was synthesized as reference (AH3, based on dihydroanthracene with ethynylbenzene side units,[21] and functionalized with thioethyltrimethylsilane). These molecules were chemisorbed on ultra flat template-stripped Au ($^{TS}$Au) surfaces used as bottom electrodes (details in the Supporting Information) to form the SAMs. Ellipsometry measurements showed that the SAMs have a thickness between 2.3 and 3.1 nm (±0.2 nm, Fig. S5 and Table S1). Compared to the geometry optimized length of the molecules (2.8-3 nm), these values suggest different molecular organizations in the SAMs. For the three molecules and whatever the protecting group, XPS measurements (S2p peak) show that the molecules are chemisorbed on the bottom electrode by Au-S bonds with the amplitude ratios of [S-Au]/[S-C] consistent with the molecule stoichiometry (Fig.S6, Table S2). UPS measurements (secondary cutoff) show that the three $^{TS}$Au/SAM samples have almost the same work function (Fig. S7). The molecular junctions (MJs) were established by gently contacting the SAMs with the tip of a conductive-atomic force microscope (C-AFM) with a loading force of ≈ 6-8 nN, Fig. 1b. This approach allows to fabricate a nanoscale MJ $^{TS}$Au-molecules//PtIr tip (top electrode area of ≈ 6-7.5 nm$^2$), *i.e.* a MJ made of ≈ 20-25 molecules (see the Supporting Information, section 3), where "-" indicates a chemical bond and "//" a mechanical contact. We acquired hundreds of DC current-voltage (I-V) curves by contacting the SAM with the C-AFM tip at different places for statistical analysis of the electron transport (ET) properties of the MJs with the laser source turned on or off (Fig. 1c). Two laser sources (continuous wave) were used, a quantum cascade laser (QCL) at 30 THz and a gas laser at 2.5 THz, with irradiating power up to 43 mW on the MJs.

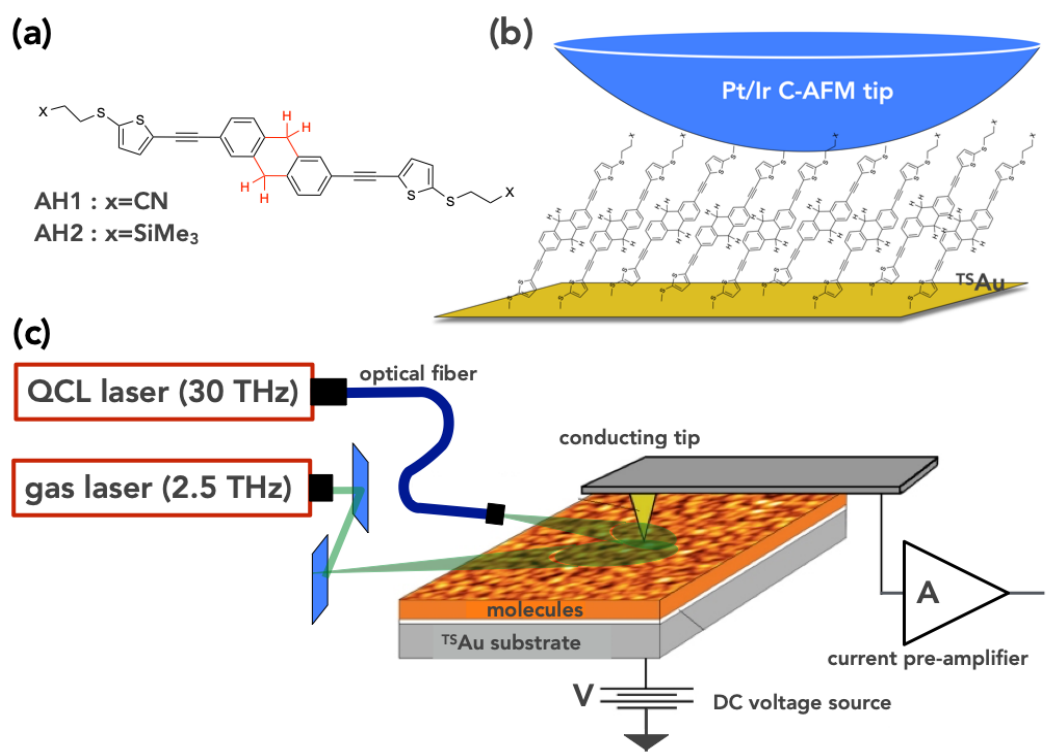

*Figure 1.* Illustration of the molecular junctions and experiments. (a) Chemical structure of the 3,3'-((((9,10-dihydroanthracene-2,6-diyl)bis(ethyne-2,1-diyl))bis(thiophene-5,2-diyl))bis(sulfanediyl))dipropanenitrile (**AH1**) and 2,6-bis((5-((2-(trimethylsilyl)ethyl)thio)thiophen-2-yl)ethynyl)-9,10-dihydroanthracene (**AH2**). (b) Scheme of the molecular junction with the SAM connected by the tip of the C-AFM. (c) Scheme of the experiment. THz waves are focalized under the tip of the C-AFM (see details in the section 3 of the Supporting Information).

All three molecules showed a marked NDC behavior at RT, which is suppressed under THz irradiations. Figure 2a shows the I-V dataset (116 I-V traces) for the $^{TS}$Au-AH1//PtIr tip MJs without THz wave irradiation. We observed a negative differential conductance (NDC), *i.e.*, a decrease in the current when increasing the applied voltage, for all the I-V traces shown in the dataset. The I-V curves are rather symmetric and the voltage positions of the peaks of current at positive and negative voltages (inset Fig. 2a), extracted from the statistical distribution given in Fig. 2c, are $V_{P+}$ = 0.42 ± 0.05 V and $V_{P-}$ = -0.48 ± 0.05 V. The peak-to-valley current ratios ($R_{PV}=I_P/I_V$) range between 1.5 and 18 with a mean value $R_{PV}$ = 3.1 ± 1.3 (Fig. 2d). When the 30 THz laser is turned on, the I-V dataset measured under irradiation no longer exhibited the NDC effect (Fig. 2b). The switching is fast and reversible. The inset in Fig. 2b presents the time evolution of the current measured at the positive peak voltage while turning on and off the 30 THz laser (5 cycles). The switching occurs in less than 24 ms, the minimal time accessible between two successive acquisitions of the current with our C-AFM equipment. The real switching time is much faster, the switching being purely electronic in nature (*e.g.* ≈2 ps in the model proposed in Ref. [20]). These NDC and THz switching behaviors are reproducible (checked on another sample of AH1 from another batch, Fig. S8).

The same behavior was observed for the $^{TS}$Au-AH2//PtIr tip MJs (Fig. S9), albeit we note that the peak voltages $V_{p+}$ and $V_{p-}$ are slightly larger (0.6-0.8 V in absolute value) than for the AH1 junction. The peak-to-valley ratios remain in the range 2-10, comparable to the AH1 junctions. This slight difference between the AH1 and AH2 MJs (while the core molecules are identical) is discussed below. This THz switching effect was also observed for SAM-based MJs made of molecules AH3 with ethynylphenyl sides (Fig. S10) for which a NDC behavior was already reported at the single molecule level using mechanically controlled break junction.[21]

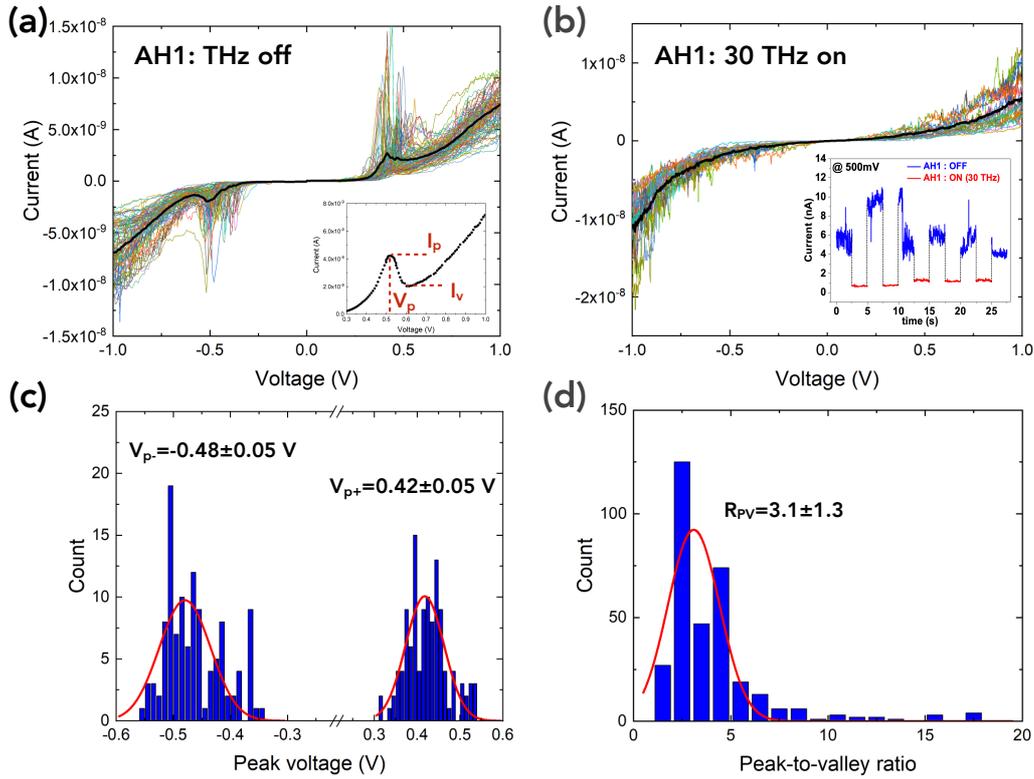

*Figure 2*. NDC and THz switching effects. (a) Current-voltage dataset for the $^{TS}$Au-AH1//PtIr tip MJs without THz wave irradiation (116 I-V curves, different colors, acquired at different places on the SAM, see the Supporting Information). The bold black line is the mean $\bar{I}$-V curve. The inset illustrates, for one of the I-V of the dataset, the determination of the peak and valley currents and the peak voltage. (b) Current-voltage (23 I-V curves, different colors) dataset for the $^{TS}$Au-AH1//PtIr tip MJs under a 30 THz continuous wave irradiation (at a power of 40 mW). The bold black line is the mean $\bar{I}$-V curve. The inset shows the time evolution recorded at the positive peak voltage when switching on and off the 30 THz waves, the tip of the C-AFM remaining at the same position on the SAM surface. (c) Histogram of the voltages of the peak currents (from the dataset in Fig. 2a) for the positive, $V_{P+}$, and negative, $V_{P-}$, voltages. (d) Histogram of the peak-to-valley ratios (from the dataset in Fig. 2a) calculated for every I-V as depicted in the inset of Fig. 2a.

The THz switching effect depends on the microwave power and frequency (Fig. 3). We have observed a power threshold. When the 30 THz power applied on the $^{TS}$Au-AH1//PtIr tip MJs is below 28 mW, we do not observe the disappearance of the NDC peaks even for a period of irradiation up to few minutes (Figs. 3a-b), a time much larger than the switching time previously observed (Fig. 2b). Between 28 and 40 mW, the NDC peaks are suppressed (Fig. 2b, Figs. 3c-d) and the current recorded under irradiation is almost similar as the "background" current of the I-V without irradiation (*i.e.*, the current superimposed to the NDC peaks, Fig. 3c). The switching effect depends also on the THz frequency. We did not observe the suppression of the NDC effect under irradiation at 2.5 THz (Figs. 3e-f). As a control experiment, we measured I-Vs under 30 THz irradiation (same power of 40 mW) of MJs made of SAMs of oligo(phenylene ethynylene) (OPE). We did not observe any modification of the I-Vs (Fig. S15). The current suppression observed for the AH1, AH2 and AH3 MJs is specific to the NDC behavior of these π-σ-π molecules. We note that we have not observed, under these THz irradiation conditions, an increase of the background current due to THz photon-assisted tunneling effect (PAT)[9, 17] see section 4.2 in the Supporting Information.

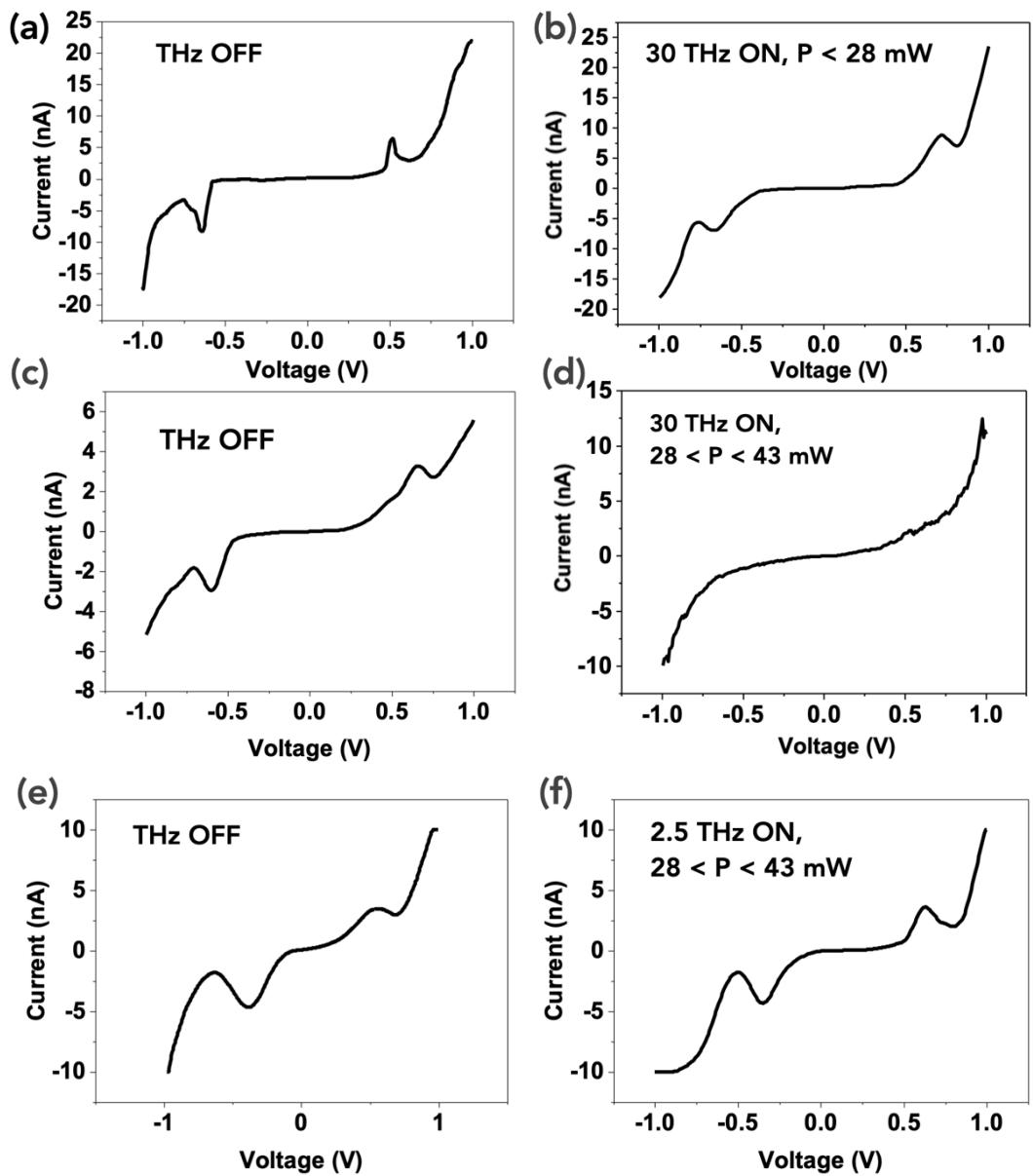

*Figure 3. THz wave power and frequency: effect on the NDC switching.* I-V curves measured on the $^{TS}$Au-AH1//PtIr tip MJ without/with THz irradiation (30 THz) at different powers (P) applied during few tens of seconds to few minutes: (a-b) P < 28 mW; (c-d) 28 < P < 43 mW. (e-f) at 2.5 THz (same P=40 mW, several minutes).

The I-V datasets with the THz laser off are well explained by the two-site model developed for this type of π-σ-π molecule.[21] In a simple picture, the peak of the current occurs when the applied voltage aligns the energy of the two π-arms giving rise to a resonant ET inside the MJ. Increasing the applied voltage, the NDC effect results from the breaking of this resonant condition, pulling the energy of the two π sites apart. The analytical model developed in Refs. [21, 22] was used and it well reproduced the NDC behavior of our MJs (Fig. 4 and Fig. S11). The model (inset in Fig. 4a) is parametrized by the energy levels $\varepsilon_0$ of the π moieties (at zero bias, with respect to the Fermi energy of the electrodes), the electronic coupling energy Γ with the electrodes and τ the coupling between the two π sites (Eq. S4 in the Supporting Information). Figure 4a shows the fit of the mean Ī-V curve of the $^{TS}$Au-AH1//PtIr tip MJ, and Figs. 4b-c the fits on two representative I-V curves of the dataset with a clear NDC behavior (the NDC effect being reduced on the mean Ī-V due to data dispersion). All the individual I-V of the dataset shown in Fig. 2a were fitted and the histograms of the fitted parameters are shown in Figs. 4d-f. The energy levels are Gaussian distributed, with $\varepsilon_0$=0.35±0.04 eV, while the coupling energies are more dispersed with 3<Γ<140 meV (with ∼ 78% of the values below 50 meV) and 2<τ<10 meV (∼ 85% of the values <5 meV). Similar results are obtained for the AH2 molecular junctions (Fig. S12). We note a slightly larger $\varepsilon_0$ for the AH2 MJs, which might be related to a slightly more densely-packed molecular organization in the AH2 SAM (see section 2.3 in the Supporting Information). The model also fits the I-V of the AH3 MJs (Fig. S13), as already reported for single molecule experiments.[21] The general observed trend that τ < Γ is qualitatively in agreement with the structure of the junction (sigma bonds *vs.* chemisorbed thiol bonds), and the *ab initio* calculations (τ ≈ 10 meV, Fig. S16 and Γ ≈ 150-200 meV, Fig. S19).

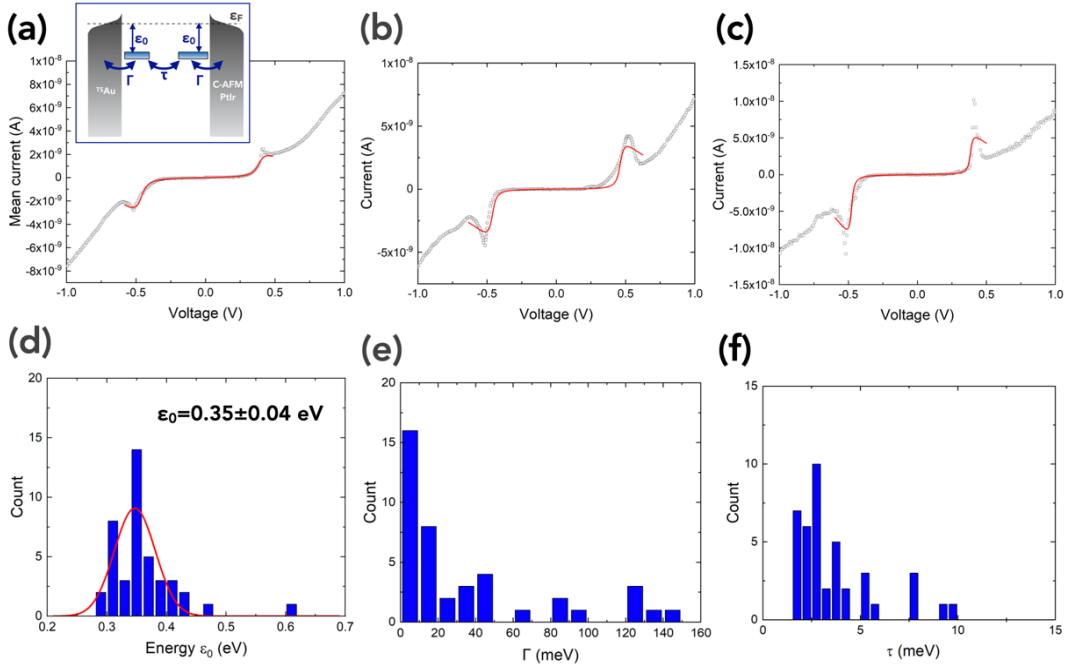

*Figure 4. Analysis with a two-site model. Fits of the two-site model (Eq. S4) on (a) the mean Ī-V of the $^{TS}$Au-AH1//PtIr tip MJ (dataset of Fig. 2a) and (b-c) two typical I-V from the dataset with a marked NDC behavior. The fit parameters are: (a) $\varepsilon_0$=0.32 eV, $\Gamma$=101 meV, $\tau$=2 meV; (b) $\varepsilon_0$=0.35 eV, $\Gamma$=35 meV, $\tau$=4 meV; (c) $\varepsilon_0$=0.33 eV, $\Gamma$=119 meV, $\tau$=3.5 meV. The fits are limited between the voltage positions of the two current valleys since the monotonous increase of the background "off-resonance" current with the voltage at higher voltages is not taken into account by this model (see Fig. 4d in Ref. [21]). (d-f) Histograms of the fit parameters obtained by fitting all individual I-V curves of the dataset (Fig. 2a): (d) energy of molecular orbital $\varepsilon_0$, the red line is the fit with a Gaussian distribution; (e) electrode coupling energy $\Gamma$ and (f) intramolecular electronic coupling energy $\tau$.*

Under the 30 THz irradiation, the I-V curves (Fig. 2b) are well fitted with a single-energy level (SEL, Eq. S6 of the Supporting Information) with $\varepsilon_{0\text{-THz}}$ = 0.50±0.09 eV and coupling energies to the electrodes in the 0.05 to 3 meV range (Fig. S14). A similar result was obtained for the AH2 MJs, $\varepsilon_{0\text{-THz}}$ = 0.56 ± 0.04 eV and coupling energies to the electrodes in the 0.01 to 1.5 meV range (Fig. S14). The meaning of this virtual molecular orbital level $\varepsilon_{0-THz}$ under dynamic excitation is discussed in below.

To explain and rationalize the experimental results, we performed DFT (density functional theory) calculations coupled to the NEGF (non-equilibrium Green function) formalism[31] (Supporting Information, section 5). Figure 5 shows the optimized geometry of the AH1 junction considering chemisorbed S-Au bonds at the two ends (Fig. 5a), the calculated transmission probability T(E) at several voltages and the localization of the HOMO and HOMO-1 orbitals (Fig. 5b). At zero bias, the HOMO and HOMO-1 are degenerate at *ca.* 0.47-0.49 eV below the Fermi energy of the electrodes. When a voltage is applied the two MOs split by following the different electrodes, with the HOMO (HOMO-1, respectively) localized on the right (left, respectively) side of the molecule and the energy moving upward (downward, respectively) – more data in Fig. S16. The NDC behavior is due to the opposite shift of the HOMO and HOMO-1 with the applied voltage, the HOMO-1 being always pulled out from the transmission energy window at V>$V_P$ ($V_P$=0.7 V in this calculation), thus decreasing T(E) and the current, confirming the NDC mechanism proposed for single molecule break-junction measurements.[21] Figure 5c shows the energy difference, Δ, between the two MOs versus the applied voltage. The simulated I-V curve is shown in Fig. 5d with NDC peaks at +/- 0.7 V.

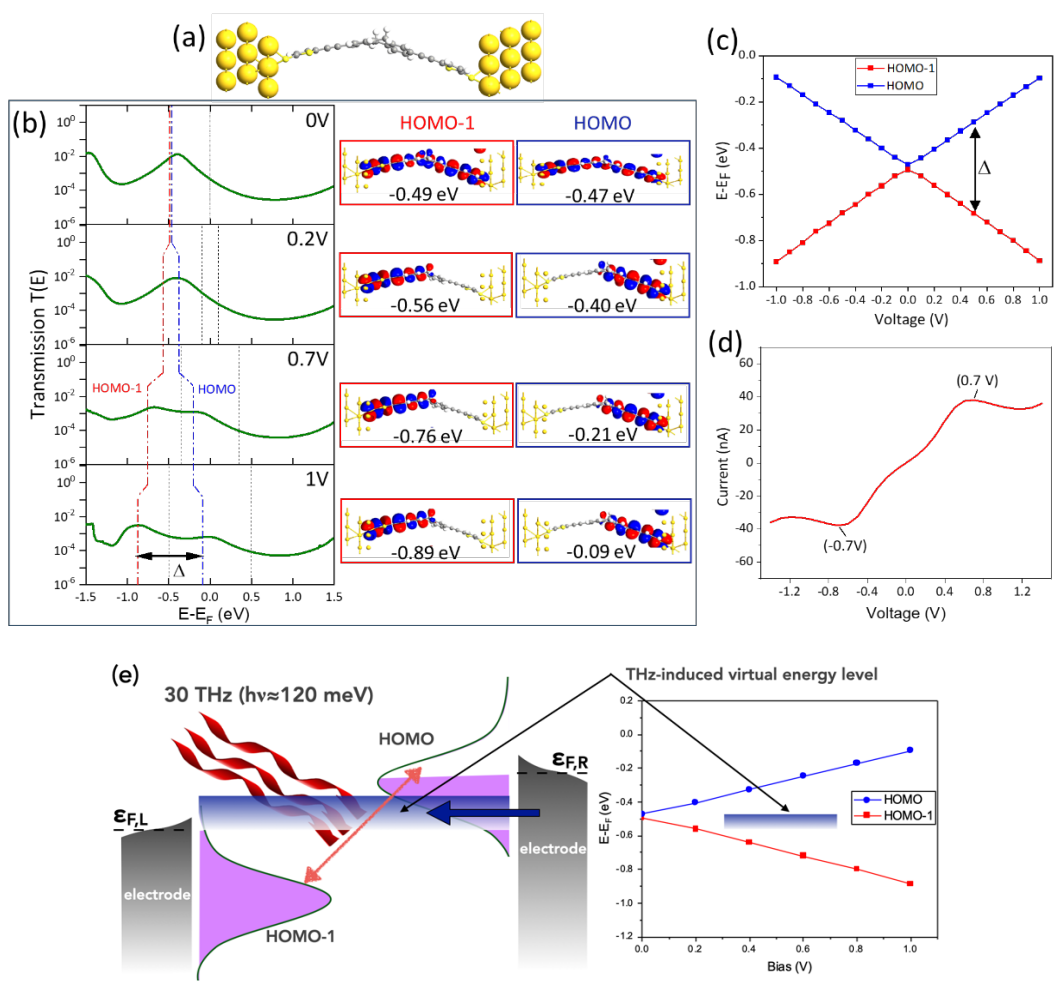

*Figure 5. DFT and NEGF simulations.* *(a) Optimized geometry of the Au-AH1-Au molecular junctions (the molecule is V-shaped with an S-to-S length of 2.2 nm) - see also Fig. S5 and Table S1 in the Supporting Information. (b) Calculated transmission coefficient, T(E), at several applied voltages. The blue and red dashed lines highlight the voltage-dependent shift of the energy position of the HOMO and HOMO-1, respectively. The vertical dark dashed lines indicate the position of the Fermi energy of the two electrodes (the energy window used to calculate the current, see the Supporting Information). The blue and red boxes show the corresponding electronic density of the HOMO and HOMO-1 orbitals, with their energy position with respect to the Fermi energy of the electrode. (c) Evolution of*

*the energy gap Δ between the HOMO and HOMO-1 with the voltage. The calculated Δ is fitted with Eq. S5 to determine the fraction of the voltage that drops inside the molecule (parameter α ≈ 0.8, see Supporting Information). Due to the symmetry of the MJ, the same results are obtained at negative voltages. (d) Simulated I-V curves with NDC peaks at +/- 0.7V. (e) Illustration of the THZ switching effect. Schematic description of the molecular junction under THz wave irradiation and at a voltage around the peak voltage $V_P$. The HOMO and HOMO-1 are split by about 0.5 eV (Fig. 5c) and broadened by the large coupling with the electrodes (Γ up to 150-200 meV as measured in the "dark", i.e. without THz, Figs. 4 and S11). The 30 THz photons (energy of 120 meV) induce charge transfers and interactions between the two levels, which can cancel the NDC peak current (see text). The ET through the MJ is described by a single virtual level that lies at an energy between the HOMO and HOMO-1 levels.*

The simulations are in good agreement with the experiments. The simulated NDC peaks are at +/-0.7V (~ +/-0.5 to +/-0.7 V in the experiments). At zero bias, the degenerate HOMO/HOMO-1 levels lie at 0.47 - 0.49 eV below the Fermi energy, while the value deduced from the fits of the two-site model on the experimental datasets gives $\varepsilon_0$ around 0.35 - 0.47 eV (*vide supra*). Since the protecting groups could remain at the top surface of the SAMs, we also consider calculations for the Au-AH1//Au MJ (one side chemisorbed, one with mechanical contact with the C-AFM tip). Globally, we end up with similar conclusions: the splitting of the HOMO and HOMO-1 levels gives rise to the NDC behavior, but due to the strong asymmetry, the NDC peak is only present at a positive voltage (Fig. S17). We conclude that in our experiments a majority of the protecting groups are spontaneously removed when the C-AFM tip is put in contact with the surface of the SAM, and that Pt-S bonds are formed like in SAMs,[32] resulting in the experimentally observed almost symmetric NDC behaviors in agreement with our calculations. No significant rectification is expected from the different nature of the electrodes ($^{TS}$Au surface and PtIr tip) since they have almost the same work function: 4.2-4.4 eV for $^{TS}$Au electrodes functionalized by the SAMs (UPS measurement, Fig. S7b in the Supporting Information) and ≈ 4.3 eV for the PtIr tip.[33]

The observed suppression of the NDC behavior under THz radiation is qualitatively in agreement with the theoretical model of Orellana and Claro,[20] where the dynamic coupling between the THz wave and the molecular orbitals of the two sites of the molecule induces a charge transfer between the MOs turning the current to a low value. This model predicts the existence of a threshold in the THz field strength to trigger the switching, as experimentally observed. A sufficiently large field (square root the power of the incoming THz wave) is required to observe the THz switch. Several authors have theoretically proposed mechanisms that can suppress the conductance in MJs under electromagnetic wave irradiation. For instance, in a molecular wire (ET through sequential tunneling between adjacent levels), a phenomenon known as "coherent destruction of tunneling"[25] was theoretically described in various ac-field driven MJs.[26-28] In a two-level system, a resonance between the two levels and a surface plasmon polariton (SPP) can induce a decrease of the current.[29, 30] In Ref. [29], the authors considered the HOMO and LUMO as the two-level system under visible light excitation, while in Ref. [30] the two-level system is made of a delocalized orbital between the electrodes of the MJs and a localized one in the middle of the molecule. Here, the 30 THz ac driven field, enhanced by the plasmonic cavity created between the C-AFM tip and the Au surface (field increased by a factor $10^2$-$10^6$, the exact value being dependent on the precise shape of the electrodes, cavity modes, nature of material of the electrode spacer, see section 4.2 in the Supporting Information)[9, 34] can be coupled with the molecules in the MJs. We propose that the THz photons with an energy of 120 meV (at 30 THz) can induce a destructive resonance (dynamic charge transfer in Ref. [20]) between the split HOMO and HOMO-1 levels. Considering the existing large broadening of the HOMO levels (as measured in the "dark" (no THz) by the large electrode coupling energy Γ, Figs. 4 and S11), these orbitals can overlap with an energy band of at least a width of 120 meV (Fig. 5e) comparable to the energy of the 30 THz photons. Thus, the THz wave can induce charge transfers and interactions between these two levels, which can cancel the NDC peak current. The current at the NDC peak is strongly reduced and only a virtual MO is generated by this ac driven ET in the MJs.[35, 36] This virtual energy level is detected at an energy $\varepsilon_{0\text{-THz}}$ when fitting the I-Vs curves measured under the THz irradiation with the SEL model (Fig. S14). This virtual level lies between the HOMO and HOMO-1, Fig. 5e ($\varepsilon_{0\text{-THz}} \approx 0.5\text{-}0.56$ eV). At 2.5 THz, the photon energy (10 meV) is too weak to induce this coherent destruction of the ET, and no suppression of the NDC is observed on the measured I-Vs.

In summary, we have demonstrated an efficient conductance switching at room temperature in MJs made of π-σ-π molecules under the irradiation of THz waves. In the following, we envision potential applications as THz detectors. Quantum-well photodetectors (QWPs) based on electron intersubband transitions have a high responsivity among several types of THz detectors based on THz-induced electronic current variations (photocurrent) but they require working at low temperatures (<77 K).[37] The best reported performance[38] is a responsivity (at *ca.* 4 THz) of 1 AW$^{-1}$ with a detection area 1.6x10$^5$ μm$^2$ (at 5 K), i.e., 6.3x10$^{-6}$ AW$^{-1}$μm$^{-2}$. Here, the current variation at the NDC peak voltage is about ΔI = I(THz off)-I(THz on) ≈ 5 nA on average (see Fig. 2, with values up to 20 nA) at a minimum power of 28 mW. With a detection area of ≈ 10 nm$^2$ (C-AFM tip contact area, see Supporting Information, Section 3.3), we get a responsivity of 1.8x10$^{-2}$ AW$^{-1}$μm$^{-2}$ (up to 7x10$^{-2}$ AW$^{-1}$μm$^{-2}$). Our MJs present a huge increase in the responsivity (at equivalent detection area) and, moreover, they are working at room temperature.

To implement such molecular THz detectors, we need integrated MJs compatible with modern nanoelectronic fabrication processes, with electrodes of large areas and transparent to THz waves. A candidate of choice is a single layer graphene (SLG) top electrode, SLG being practically transparent to THz waves.[39] The technology is already available to deposit (soft process by transfer printing) and pattern SLG top electrodes (typically 10-100 μm in diameter) on SAMs[40] and thin molecular films[41] with low contact resistance (<1 Ω) and fabrication yield of 50-90%. The frequency response (here 30 THz) can be modulated by molecular engineering. Modifying or functionalizing the π moieties,[22, 42] the HOMO and HOMO-1 energy levels can be adapted to match with the THz photon energy, thus modulating the frequency response of the molecular detector.

## Associated Content

The Supporting Information is available free of charge at https://pubs.acs.org/doi/…..

> Details on the synthesis of the molecules, NMR characterizations, bottom electrode fabrication, thickness measurements, XPS and UPS characterizations, Conductive-AFM protocol, data analysis, theoretical models and methods, and additional results.

## Author Information

*Corresponding authors*

**Imen Hnid** – *Institute for Electronics Microelectronics and Nanotechnology (IEMN), CNRS, University of Lille, Av. Poincaré, Villeneuve d'Ascq, France; orcid.org/0000-0002-4468-5334;* E-mail: imen.hnid@iemn.fr

**Dominique Vuillaume** – *Institute for Electronics Microelectronics and Nanotechnology (IEMN), CNRS, University of Lille, Av. Poincaré, Villeneuve d'Ascq, France; orcid.org/0000-0002-3362-1669;* E-mail : dominique.vuillaume@iemn.fr

*Authors*

**Ali Yassin** – *MOLTECH-Anjou, CNRS, University of Angers, SFR MATRIX, F-49000 Angers, France; and Natural Sciences Department, School of Arts and Sciences, Lebanese American University, Lebanon; orcid.org/ 0000-0002-3047-9863*

**Imane Arbouch** – *Laboratory for Chemistry of Novel Materials, University of Mons, Belgium; orcid.org/ 0000-0002-4593-0098*

**David Guérin** – *Institute for Electronics Microelectronics and Nanotechnology (IEMN), CNRS, University of Lille, Av. Poincaré, Villeneuve d'Ascq, France; orcid.org/ 0000-0002-4338-1742*

**Colin van Dyck** – *Theoretical Chemical Physics group, University of Mons, Mons, Belgium: orcid.org/ 0000-0003-2853-3821*

**Lionel Sanguinet** – *MOLTECH-Anjou, CNRS, University of Angers, SFR MATRIX, F-49000 Angers, France; orcid.org/0000-0002-4334-9937*

**Stéphane Lenfant** – *Institute for Electronics Microelectronics and Nanotechnology (IEMN), CNRS, University of Lille, Av. Poincaré, Villeneuve d'Ascq, France; orcid.org/0000-0002-6857-8752*

**Jérôme Cornil** – *Laboratory for Chemistry of Novel Materials, University of Mons, Belgium; orcid.org/0000-0002-5479-4227*

**Philippe Blanchard** – *MOLTECH-Anjou, CNRS, University of Angers, SFR MATRIX, F-49000 Angers, France; orcid.org/0000-0002-9408-8108*

*Author Contributions*

I.H. and D.G. fabricated and characterized the SAMs. A.Y. synthesized the molecules. A.Y., L.S. and P.B. discussed synthetic procedures and performed physico-chemical and spectroscopic analysis. I.H. did all the C-AFM measurements. I.A., C.v.D. and J.C. conducted the theoretical simulations. I.H. and D.V. analyzed the I-V datasets. D.V. conceived the project, S.L. and D.V. supervised it. The manuscript was written by D.V. with the contributions and comments of all the authors. All authors have approved the final version of the manuscript.

*Note*

The authors declare no competing financial interest.

## Acknowledgements.

We acknowledge support from the ANR (projet #ANR-20-CE30-0002). We thank X. Wallart (IEMN) for the XPS and UPS measurements, J.F. Lampin (IEMN) for the loan of the THz lasers and help for their use. The work in Mons has been funded by the Fund for Scientific Research (FRS) of FNRS within the Consortium des Equipements de Calcul Intensif (CECI) under grant 2.5020.11, and by the Walloon Region (ZENOBE Tier-1 supercomputer) under grant 1117545. J.C. is an FNRS research director.

**ToC graphic.**

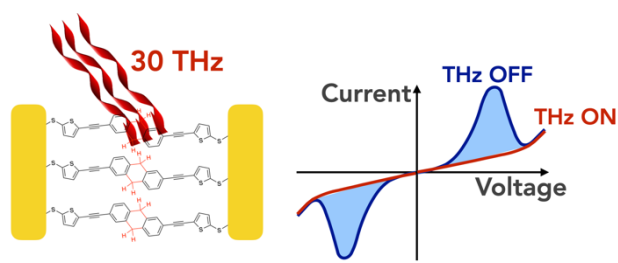

# *Supporting Information*

## Molecular Junctions for Terahertz Switches and Detectors.

Imen Hnid,[1*] Ali Yassin,[2,3] Imane Arbouch,[4] David Guérin,[1] Colin van Dyck,[5]

Lionel Sanguinet,[2] Stéphane Lenfant,[1] Jérôme Cornil,[4]

Philippe Blanchard[2] and Dominique Vuillaume.[1*]

*1. Institute for Electronics Microelectronics and Nanotechnology (IEMN), CNRS, University of Lille, Av. Poincaré, Villeneuve d'Ascq, France.*
*2. MOLTECH-Anjou, CNRS, University of Angers, SFR MATRIX, F-49000 Angers, France.*
*3. Natural Sciences Department, School of Arts and Sciences, Lebanese American University, Beirut, Lebanon.*
*4. Laboratory for Chemistry of Novel Materials, University of Mons, Belgium.*
*5. Theoretical Chemical Physics group, University of Mons, Mons, Belgium.*

# 1. Synthesis of the molecules.

## *1.1. General information*

The synthetic step to obtain each of the three molecules is based on a palladium-catalyzed Sonogashira coupling with key compound **3** (See Scheme). For that purpose synthetic blocks end-capped with different protected thiols, as the anchoring unit on gold electrodes, were iodinated to generate derivatives **4**, **5** and **6**. In parallel, 2,6-diethynyl-9,10-dihydroanthracene **3** was prepared *via* reduction of commercially available 2,6-dibromo-9,10-anthraquinone, providing intermediate compound **1**, followed by its coupling to trimethyl((tributylstannyl)ethynyl)silane under Stille reaction conditions to produce compound **2**. Careful deprotection performed with cesium carbonate generated our target compound **3**. The Sonogashira coupling conditions were optimized using the base, triethylamine, as a solvent in presence of Pd(PPh$_3$)$_2$Cl$_2$ as catalyst and CuI as cocatalyst. Very similar yields were also achieved when DMF was added in order to homogenize the reaction mixture, though it was usually avoided for purification purposes (*vide infra* the detailed synthetic procedures with NMR characterization - Figs. S1 to S3 - and High-resolution Mass Spectrometry). Having identical π-σ-π system, compounds AH1 and AH2 show similar UV-Vis absorption spectra with maximum absorption bands between 320 and 340 nm with a slight broadening of the absorption edge for AH2 (Fig. S4). When thiophene is replaced by the more aromatic phenyl moiety, the absorption maxima in AH3 shift to higher energies (311, 318, and 330 nm) consistent with a lesser conjugated system.

Reagents and chemicals from commercial sources were used without further purification. Solvents were freshly dried and purified by distillation from sodium–benzophenone ketyl. Flash chromatography was performed with analytical-grade solvents using Aldrich silica gel (technical grade, pore size 60 Å, 230-400 mesh particle size). Flexible plates ALUGRAM® Xtra SIL G UV254 from MACHEREY-NAGEL were used for TLC. Compounds were detected by UV irradiation (Bioblock Scientific). Microwave assisted reactions were performed in the cavity of a Biotage Initiator+ system in sealed reactors. Compounds purified with recycling preparative size exclusion HPLC were solubilized in HPLC grade chloroform. Before injection, the solution was filtered through a 0.45 µm PTFE filter (VWR 25 mm syringe filter w/ 0.45 µm membrane). Purification was performed on a LC-9160 NEXT system from the Japan Analytical Industry Co., Ltd. (JAI) equipped with coupled UV-Vis 4Ch NEXT and RI-700 II detectors at room temperature through a set of two JAIGEL-2H and 2.5H columns at an elution rate of 10 mL.min$^{-1}$. NMR spectra were recorded with a Bruker AVANCE III 300 ($^1$H, 300 MHz and $^{13}$C, 75 MHz) or a Bruker AVANCE DRX500 ($^1$H, 500 MHz; $^{13}$C, 125 MHz). Chemical shifts are given in ppm relative to TMS and coupling constants J in Hz. UV-Vis spectra were recorded with a Perkin Elmer 950 spectrometer. Matrix Assisted Laser Desorption/Ionization was performed on MALDI-TOF MS BIFLEX III Bruker Daltonics spectrometer. High-resolution mass spectrometry (HRMS) was performed with a JEOL JMS-700 B/E or a JEOL Spiral-TOF JMS3000.

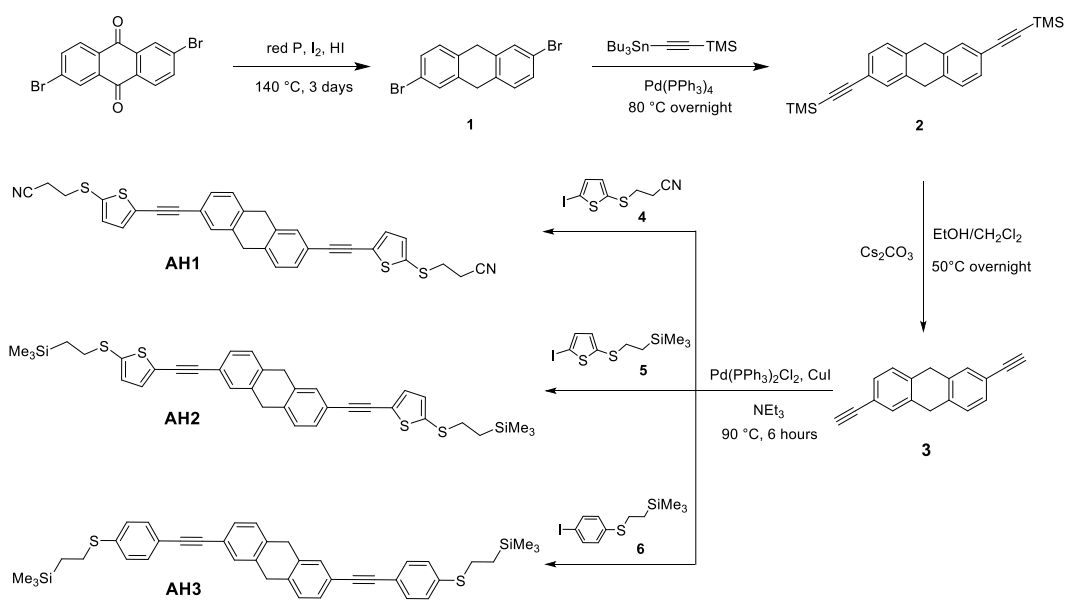

*Scheme of the synthesis routes*

## 1.2. Organic Synthesis and Characterization

**2,6-dibromo-9,10-dihydroanthracene (1):**[1]

This compound was prepared according to a previously reported procedure: 2,6-Dibromo-9,10-anthraquinone (1.32 g, 3.61 mmol), red phosphorus (1.05 g, 34 mmol) and iodine (0.25 g, 1.0 mmol) were placed in a Teflon lined hydrothermal autoclave reactor, and thoroughly mixed together with a spatula. Hydroiodic acid (13 mL, 57% w/w in water) was added and the reactor was sealed and heated at 140°C for 3 days. After cooling to room temperature, the suspension was poured into water and filtered off. The residue was washed twice with 50 mL of cold water and hot water consecutively, then twice with cold acetone. The resulting solid was dissolved and recrystallized from hot ethanol affording 0.5 g of a crystalline white solid (1.48 mmol, 41%). 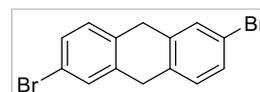 **$^1$H NMR** (300 MHz, CDCl$_3$): δ 7.43 (d, J = 2, 2H), 7.32 (dd, J = 8.0, 2.0, 2H), 7.14 (d, J = 8.0, 2H), 3.84 (s, 4H). **$^{13}$C NMR** (125 MHz, CDCl$_3$): δ 183.20, 156.64, 150.45, 150.38, 149.82, 148.73, 144.58, 144.54, 139.88, 135.74, 131.44, 131.00,

130.92, 127.59, 120.36, 120.12, 118.27, 114.92, 110.02, 55.52. **HRMS** (EI) m/z: Calc for $C_{14}H_{10}Br_2$ (M$^+$), 335.9149; Found, 335.9140.

**2,6-bis((trimethylsilyl)ethynyl)-9,10-dihydroanthracene (2):**

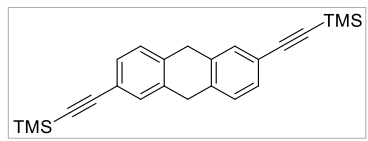

Trimethyl((tributylstannyl)ethynyl)silane (505 mg, 1.30 mmol) was added to a solution of 2,6-dibromo-9,10-dihydroanthracene (**1**) (200 mg, 0.6 mmol) and Pd(PPh$_3$)$_4$ (68 mg, 10 mol%) in 20 mL of freshly distilled dry toluene under Argon atmosphere. The mixture was heated under inert atmosphere at 90°C overnight, before being cooled to room temp. The reaction was quenched with water, extraction was performed three times with dichloromethane, and the combined organic layers were dried over MgSO$_4$. After removal of the solvent, the residue was purified by chromatography on silica gel (petroleum ether as eluent) to give the product as off-white crystalline powder (220 mg, 99% yield). **$^1$H NMR** (300 MHz, CDCl$_3$): δ 7.40 (d, J = 1.5, 2H), 7.32 (dd, J = 7.8, 1.5, 2H), 7.14 (d, J = 7.7, 2H), 3.88 (s, 4H), 0.24 (s, 18H). **HRMS** (EI) m/z: Calc for $C_{24}H_{28}Si_2$ (M$^+$), 372.1730; Found, 372.1766.

**2,6-diethynyl-9,10-dihydroanthracene (3):**

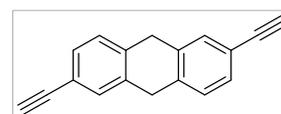

2,6-bis((trimethylsilyl)ethynyl)-9,10-dihydroanthracene (**2**) (170 mg, 0.46 mmol), caesium carbonate (445 mg, 1.37 mmol), ethanol (3 mL), and dichloromethane (3 mL) were added to a round bottom flask equipped with a stirring bar. The reaction was stirred at 50°C overnight, after which the mixture was diluted with dichloromethane and washed 3 times with water.[2] The organic layer was dried over anhydrous Na$_2$SO$_4$ and evaporated in vacuo. The crude compound was passed through a chromatography column (silica gel; eluent: petroleum ether) to give 90 mg (87% yield) of the desired crystalline yellowish solid. **$^1$H NMR** (300 MHz, CDCl$_3$): δ 7.43 (d, J = 1. 5, 2H), 7.34 (dd, J = 7.9, 1.5, 2H), 7.24 (d, J = 7.8, 2H), 3.90 (s, 4H), 3.04 (s, 2H). **HRMS** (EI) m/z: Calc for $C_{18}H_{12}$ (M$^+$), 228.0939; Found, 228.1332.

**3-(thiophen-2-ylthio)propanenitrile:[3]**

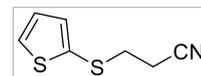

Et$_3$N (10 mg, 5 mol%) was added to a solution of thiophene-2-thiol (233 mg, 2 mmol) and acrylonitrile (106 mg, 2 mmol) in THF (1 mL) at 0°C. The resulting reaction mixture was stirred overnight at room temperature. After removing the solvent under reduced pressure, the crude residue was purified by column chromatography (silica gel; eluent: 2:3 dichloromethane/petroleum ether) to give 260 mg (1.54 mmol, 77% yield) of a brownish oil. **$^1$H NMR** (300 MHz, CDCl$_3$): δ 7.43 (dd, J = 5.4, 1.2, 1H), 7.22 (dd, J = 3.6, 1.2, 1H), 7.02 (dd, J = 5.4, 3.6, 1H), 2.97 (t, 2H), 2.60 (t, 2H). **$^{13}$C NMR** (125 MHz, CDCl$_3$): δ 135.8, 131, 130.9, 127.9, 117.8, 33.7, 18.2. **HRMS** (EI) m/z: Calc for $C_7H_7NS_2$ (M$^+$), 169.0020; Found, 169.0042.

**3-((5-iodothiophen-2-yl)thio)propanenitrile (4):**

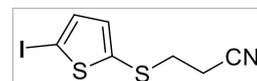

3-(thiophen-2-ylthio)propanenitrile (260 mg, 1.54 mmol) was dissolved in EtOH (10 mL) at room temperature. N-Iodosuccinimide (370 mg, 1.64 mmol) was later added followed by *p*-toluenesulfonic acid (10% mol) and the mixture was left to stir for 2 hours.[4] The reaction was quenched with a saturated solution of Na$_2$S$_2$O$_3$. The product was extracted with EtOAc and was washed with 1 M Na$_2$CO$_3$ solution, and the combined organic layers dried over MgSO$_4$. The crude compound was flash chromatographed on silica gel to yield a pale-yellow oil (450 mg, 99% yield) which was directly used in the next synthetic step. **$^1$H NMR** (300 MHz, CDCl$_3$): δ 7.18 (d, J = 3.7, 1H), 6.90 (d, J = 3.7, 1H), 2.95 (t, 2H), 2.61 (t, 2H). **HRMS** (EI) m/z: Calc for $C_7H_6INS_2$ (M$^+$), 294.8986; Found, 294.8989.

### 3,3'-(((((9,10-dihydroanthracene-2,6-diyl)bis(ethyne-2,1-diyl))bis(thiophene-5,2-diyl))bis(sulfanediyl))dipropanenitrile (AH1):

A mixture of 2,6-diethynyl-9,10-dihydroanthracene (**3**) (50 mg, 0.22 mmol), 3-((5-iodothiophen-2-yl)thio) propanenitrile (**4**) (142 mg, 0.48 mmol), bis(triphenylphosphine) palladium(II) dichloride (16 mg, 0.02 mmol), and copper(I) iodide (4.5 mg, 0.025 mmol) was degassed several times under argon atmosphere, before adding 4 mL of freshly distilled dry NEt$_3$. The mixture was heated at 90°C for 6 hours, after which the solvent was removed by rotary evaporation and the residue was diluted with water and extracted with DCM. The combined organic layers were dried over anhydrous Na$_2$SO$_4$ evaporated in vacuo, and the product was purified by column chromatography (silica gel; eluent: 2:3 dichloromethane/petroleum ether) to give 60 mg (0.1 mmol, 49% yield) of a yellow solid. **$^1$H NMR** (500 MHz, CDCl$_3$): δ 7.45 (d, J = 1.5, 2H), 7.36 (dd, J = 7.7, 1.5, 2H), 7.28 (d, J = 7.7, 2H), 7.15 (d, J = 3.7, 2H), 7.11 (d, J = 3.7, 2H), 3.01 (s, 4H), 3.01 (t, 4H), 2.64 (t, 4H). **$^{13}$C NMR** (125 MHz, CDCl$_3$): δ 137.28, 136.47, 135.57, 132.46, 132.34, 130.46, 129.60, 128.70, 127.71, 120.37, 117.71, 94.81, 81.42, 35.94, 33.81, 18.35. **HRMS** (EI) m/z: Calc for C$_{32}$H$_{22}$N$_2$S$_4$ (M$^+$), 562.0666; Found, 562.0771

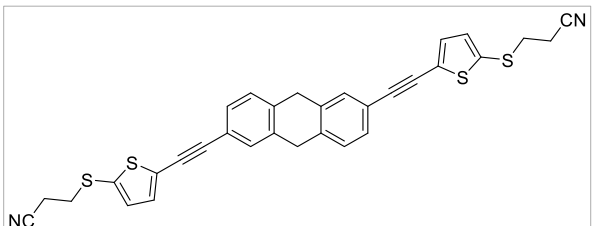

### (2-bromoethyl)trimethylsilane:[5]

Under argon atmosphere, a solution of phosphorus tribromide (0.4 mL, 2.2 mmol) in anhydrous dichloromethane was added dropwise to a mixture of vinyltrimethylsilane (1 mL, 6.8 mmol) and silica gel (3 g) in anhydrous CH$_2$Cl$_2$ cooled to -10°C. After 10 min of stirring at this temperature, the reaction mixture was allowed to warm to 20°C over a period of 0.5 h, after which silica gel was separated by filtration. The filtrated solution was washed with a saturated aqueous solution of Na$_2$CO$_3$, dried over Na$_2$SO$_4$, and concentrated to dryness to give pure (2-bromoethyl)trimethylsilane as slightly yellow oil (600 mg, 50% yield). The compound is not stable at room temperature, even when stored at +4°C, hence was directly used after synthesis. **$^1$H NMR** (300 MHz, CDCl$_3$): δ 3.59– 3.55 (m, 2H), 1.39–1.36 (m, 2H), 0.04 ppm (s, 9H).

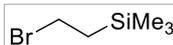

**2-((5-iodothiophen-2-yl)thio)ethyl trimethylsilane (5)**:

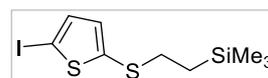

Under Argon atmosphere, a solution of CsOH.H$_2$O (205 mg, 1.22 mmol) in dry MeOH (2 mL) was added dropwise to a solution of 3-((5-iodothiophen-2-yl)thio)propanenitrile (**4**) (300 mg, 1.02 mmol) in 5 mL of dry DMF. The reaction mixture was stirred for 1 h at room temperature before addition of a solution of (2-bromoethyl)trimethylsilane (221 mg, 1.22 mmol) in 2 mL dry DMF. After 4 hours of additional stirring at 20°C and evaporation of the solvents, the residue was dissolved in CH$_2$Cl$_2$ and the organic phase was washed with water, dried over MgSO$_4$, and concentrated under reduced pressure. Purification by chromatography on silica gel (petroleum ether as eluent) gave the product as a dark yellow oil (230 mg, 66% yield). **$^1$H NMR** (500 MHz, CDCl$_3$): δ 7.12 (d, J = 3.7, 1H), 6.78 (d, J = 3.7, 1H), 2.82 (t, 2H), 0.89 (t, 2H), 0.01 (s, 9H). **$^{13}$C NMR** (125 MHz, CDCl$_3$): δ 141.1, 137.9, 135.3, 75.7, 35.9, 18.0, -1.3. **HRMS** (EI) m/z: Calc for C$_7$H$_7$NS$_2$ (M$^+$), 169.0020; Found, 169.0042.

**2,6-bis((5-((2-(trimethylsilyl)ethyl)thio)thiophen-2-yl)ethynyl)-9,10-dihydro anthracene (AH2)**:

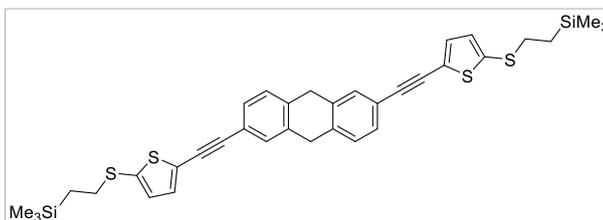

A mixture of 2,6-diethynyl-9,10-dihydroanthracene (**3**) (60 mg, 0.26 mmol), 2-((5-iodothiophen-2-yl)thio)ethyltrimethylsilane (**5**) (185 mg, 0.54 mmol), bis(triphenylphosphine) palladium(II) dichloride (19 mg, 0.026 mmol), and copper(I) iodide (6 mg, 0.03 mmol) was degassed several times under argon atmosphere, before adding 4 mL of freshly distilled dry NEt$_3$. The mixture was heated under inert atmosphere at 90°C for 6 hours, after which the solvent was removed by rotary evaporation and the residue was diluted with water and extracted with dichloromethane. The combined organic layers were dried over anhydrous Na$_2$SO$_4$, evaporated in vacuo and the product was flashed through a short plug of silica to remove the catalyst and baseline impurities. Then, to reach a high degree of purity, the resulting product was injected in a recycling preparative HPLC following the procedure described above to give 100 mg (0.15 mmol, 58% yield) of a crystalline yellow solid. **$^1$H NMR** (500 MHz, Acetone-d$_6$): δ 7.51 (d, J = 1.3, 2H), 7.41-7.37 (m, 4H), 7.23 (d, J = 3.7, 2H), 7.08 (d, J = 3.7, 2H), 4.02 (s, 4H), 2.99–2.96 (m, 4H), 0.98–0.94 (m, 4H), 0.04 (s, 18H). **$^{13}$C NMR** (125 MHz, Acetone-d$_6$): δ 139.09, 137.51, 137.25, 131.96, 130.41, 129.53, 127.87, 127.80, 121.22, 120.08, 89.72, 88.69, 35.52, 28.32, 16.51, -2.36. **HRMS** (EI) m/z: Calc for C$_{40}$H$_{44}$S$_2$Si$_2$ (M$^+$), 644.2423; Found, 644.2263.

**2-((4-bromophenyl)thio)ethyl trimethylsilane**:[6]

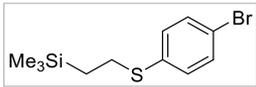

An oven-dried microwave tube containing a stirring bar was charged with 4-bromothiophenol (1.1 g, 5.9 mmol), vinyltrimethylsilane (0.7 g, 1.07 mL, 7.06 mmol) and di-tert-butyl peroxide (0.13 g, 0.88 mmol). A septum-cap was crimped on the tube to form a seal. The mixture was degassed 3 times by vacuum-argon filling cycles before exposing the reaction to microwave irradiation for 2 hours at 100°C. After cooling down to room temperature, the reaction mixture was diluted with hexane and washed twice with aqueous NaOH solution (1 m). The organic layer was dried over MgSO$_4$ and subsequently concentrated under reduced pressure. Finally, the crude product was purified by column chromatography (silica gel, petroleum ether as eluent) to yield the desired product as a colorless liquid (1.2 g, 71% yield). **$^1$H NMR** (300 MHz,

CDCl$_3$): δ 7.37 (d, J = 8.6, 2H), 7.14 (d, J = 8.6 Hz, 2H), 2.91 (m, 2H), 0.90 (m, 2 H), 0.02 (s, 9 H). **$^{13}$C NMR** (75 MHz, CDCl$_3$): δ 136.7, 132, 130.6, 119.6, 29.9, 17, –1.56. **HRMS** (EI) m/z: Calc for C$_{11}$H$_{17}$BrSSi (M$^+$), 288.0004; Found, 288.0245.

**2-((4-iodophenyl)thio)ethyl trimethylsilane (6)**:[7]

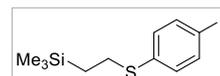

The bromo derivative (1.0 g, 3.46 mmol) was dissolved in anhydrous THF (30 mL) in a Schlenk tube under argon, cooled to -78 °C and degassed. *tert*-BuLi (1.7 M in pentane, 3.66 mL, 6.22 mmol) was added dropwise over a period of 15 min and the resulting yellow solution was stirred at -78 °C for 45 min. Iodine dissolved in THF (20 mL) in a second Schlenk flask under argon and cooled to -78 °C was cannulated dropwise into the mixture. The dark solution was stirred at -78 °C for 30 min, then allowed to warm to room temperature and stirred for an additional 2 hours. The mixture was quenched with Na$_2$SO$_3$, and the aqueous layer was extracted with DCM three times. The combined organic layers were subsequently washed with brine and dried over MgSO$_4$. The end residue was purified using column chromatography on silica gel with petroleum ether as eluent to provide the product (0.8 g, 70% yield) as a pale yellow oil. **$^1$H NMR** (500 MHz, CDCl$_3$): δ 7.58 (d, J = 8.5, 2H), 7.03 (d, J = 8.5, 2H), 2.94-2.91 (m, 2H), 0.92-0.89 (s, 9H). **$^{13}$C NMR** (125 MHz, CDCl$_3$): δ 137.9, 137.5, 130.1, 90.4, 29.5, 17, -1.53. **HRMS** (EI) m/z: Calc for C$_{11}$H$_{17}$ISSi (M$^+$), 335.9865; Found, 335.9842.

**2,6-bis((4-((2-(trimethylsilyl)ethyl)thio)phenyl)ethynyl)-9,10-dihydroanthracene (AH3)**

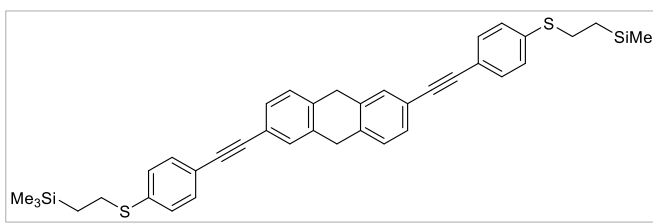

A mixture of 2,6-diethynyl-9,10-dihydroanthracene (**3**) (60 mg, 0.26 mmol), 2-((4-iodophenyl)thio)ethyl trimethylsilane (**6**) (195 mg, 0.58 mmol), bis(triphenylphosphine) palladium(II) dichloride (19 mg, 0.026 mmol), and copper(I) iodide (6 mg, 0.03 mmol) was degassed several times under argon atmosphere, before adding 4 mL of freshly distilled dry NEt$_3$. The mixture was heated under inert atmosphere at 90°C for 6 hours, after which the solvent was removed by rotary evaporation and the residue was diluted with water and extracted with dichloromethane. The combined organic layers were dried over anhydrous Na$_2$SO$_4$, evaporated in vacuo and the product was flashed through a short plug of silica to remove the catalyst and baseline impurities. Then, to reach a high degree of purity, the resulting product was injected in a recycling preparative HPLC following the procedure described above to give 90 mg (0.14 mmol, 53% yield) of a crystalline white solid. **$^1$H NMR** (500 MHz, Acetone-d$_6$): δ 7.51 (s, 2H), 7.46 (m, 4H), 7.38 (s, 4H), 7.33 (m, 4H), 4.00 (s, 4H), 3.10–3.06 (m, 4H), 0.98–0.94 (m, 4H), 0.08 (s, 18H). **$^{13}$C NMR** (125 MHz, Acetone-d$_6$): δ 139.09, 137.51, 137.25, 131.96, 130.41, 129.53, 127.87, 127.80, 121.22, 120.08, 89.72, 88.69, 35.52, 28.32, 16.51, -2.36. **HRMS** (EI) m/z: Calc for C$_{40}$H$_{44}$S$_2$Si$_2$ (M$^+$), 644.2423; Found, 644.2

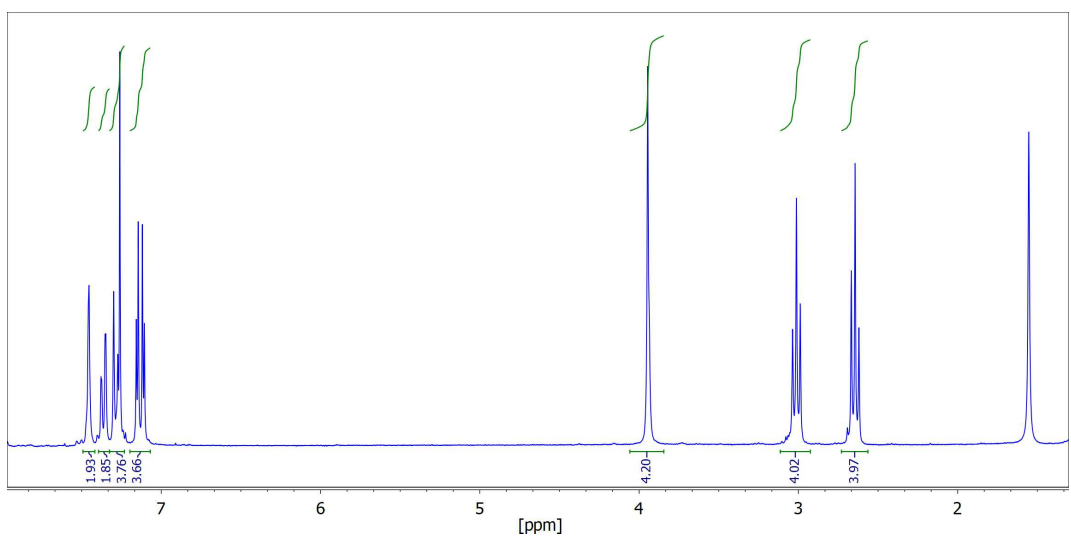

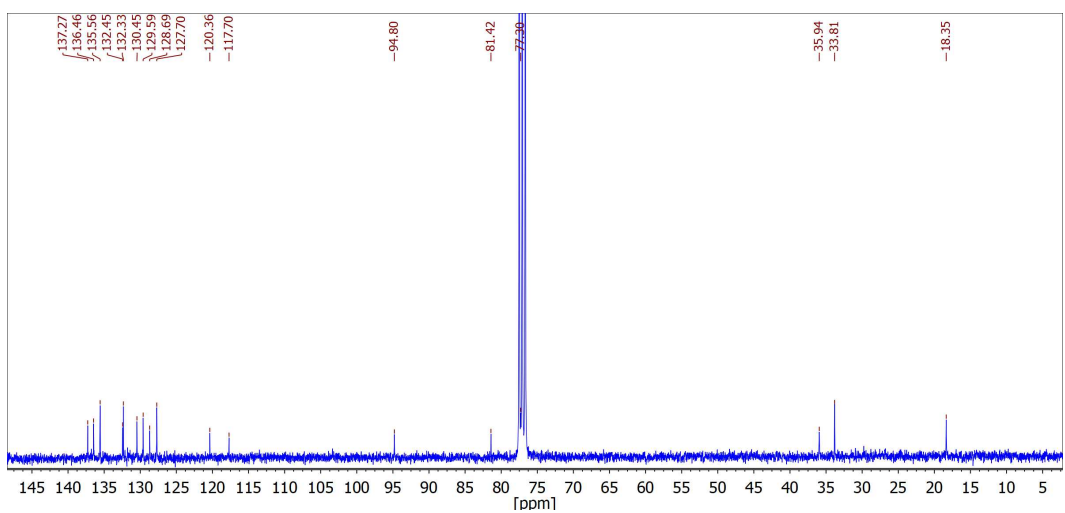

*Figure S1*. $^1$H (top, 500 MHz) and $^{13}$C (bottom, 125 MHz) NMR spectra of AH1 in CDCl$_3$ recorded at 20 °C.

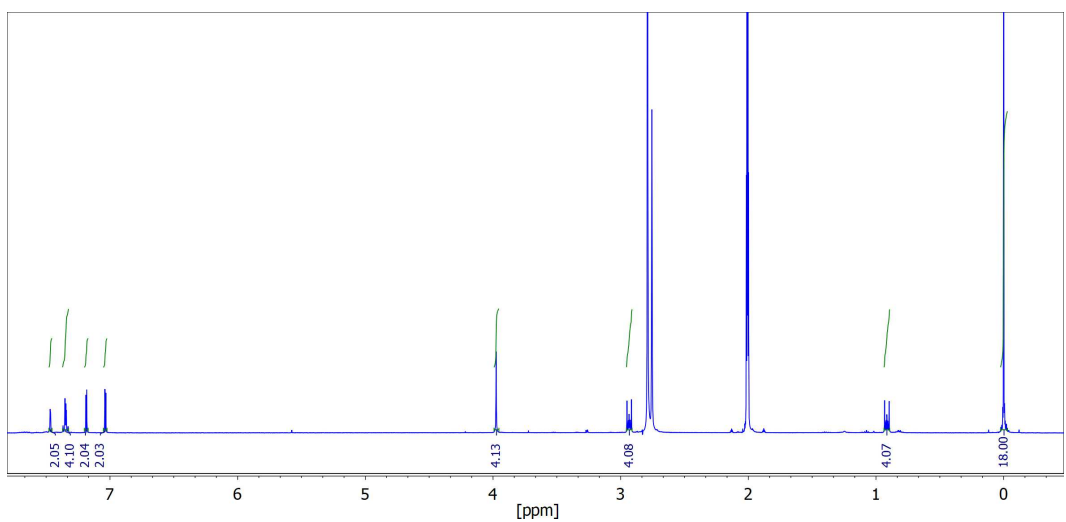

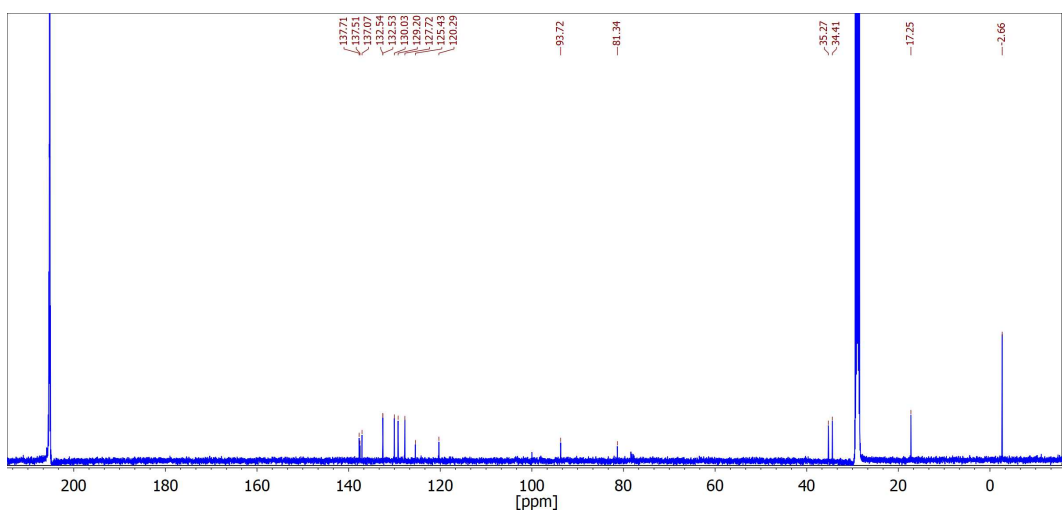

**Figure S2.** $^1$H (top, 500 MHz) and $^{13}$C (bottom, 125 MHz) NMR spectra of AH2 in $(CD_3)_2CO$ recorded at 20 °C.

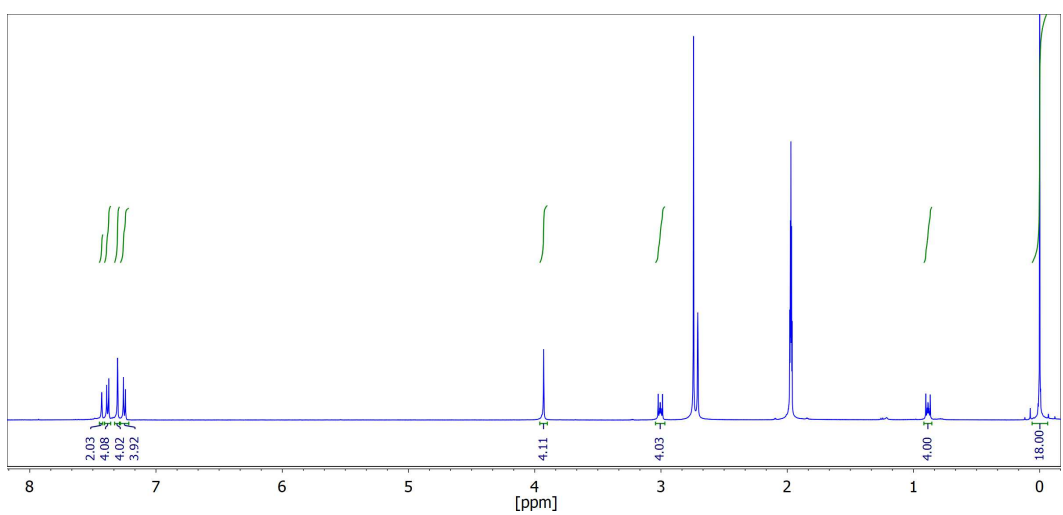

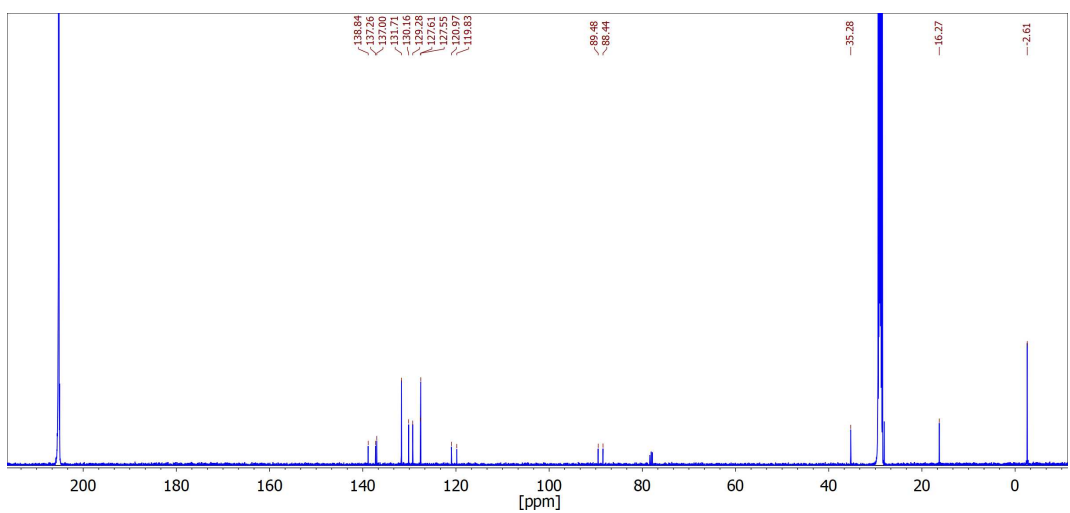

**Figure S3.** $^1$H (top, 500 MHz) and $^{13}$C (bottom, 125 MHz) NMR spectra of AH3 in $(CD_3)_2CO$ recorded at 20 °C.

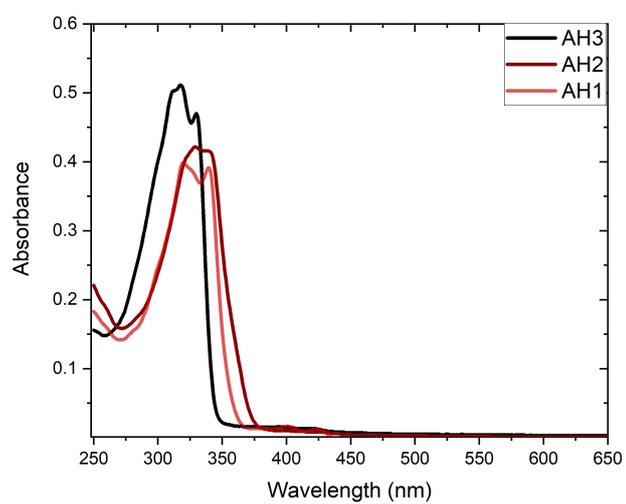

**Figure S4**. UV-Vis absorption spectra of the molecules in $CHCl_3$ recorded at 20 °C.

## 2. Molecular junction fabrication and characterization.

### 2.1. Template stripped bottom electrode

Ultraflat template-stripped gold surfaces ($^{TS}$Au), with rms roughness of ≈0.4 nm were prepared according to the methods previously reported.[8-10] In brief, a 300–500 nm thick Au film was evaporated on a very flat silicon wafer covered by its native $SiO_2$ (rms roughness of ≈0.4 nm), which was previously carefully cleaned by piranha solution (30 min in 7:3 $H_2SO_4/H_2O_2$ (v/v); **Caution**: Piranha solution is a strong oxidizer and reacts exothermically with organics), rinsed with deionized (DI) water, and dried under a stream of nitrogen. Clean 10x10 mm pieces of glass slide (ultrasonicated in acetone for 5 min, ultrasonicated in 2-propanol for 5 min, and UV irradiated in ozone for 10 min) were glued on the evaporated Au film (UV-polymerizable glue, NOA61 from Epotecny), then mechanically peeled off providing the $^{TS}$Au film attached on the glass side (Au film is cut with a razor blade around the glass piece). Our bottom $^{TS}$Au electrodes have a rms roughness of 0.4 nm.[11]

### 2.2. Self-assembled monolayers (SAMs)

For the three molecules, the SAMs were formed in solution. Under a nitrogen atmosphere, we exposed the freshly gold surfaces to $10^{-3}$ M solution of the molecules in tetrahydrofuran (THF)/ethanol (1/8 v/v) during 20 h. The deprotection is expected via direct reaction with the Au surface.[12] Then, we rinsed the treated substrates with THF followed by a cleaning in an ultrasonic bath of THF/ethanol (1/1 v/v) during 1 min and then dried with $N_2$ (see details in Supplementary Information Section 2).

### *2.3. Thickness measurements of the SAMs*

The thickness of the SAMs was measured using a UVISEL (Horiba Jobin Yvon) spectroscopic ellipsometer equipped with DeltaPsi 2 data analysis software. The system acquired a spectrum ranging from 2 to 4.5 eV (300–750 nm) with intervals of 0.1 eV (or 15 nm). The data were taken at an angle of incidence of 70°, and the compensator was set at 45°. We fit the data by a regression analysis to a film-on-substrate model as described by their thickness and their complex refractive indexes. We estimated the accuracy of the SAM thickness measurements at ± 2 Å.[13] First, a background for the substrate before monolayer deposition was recorded. We acquired three reference spectra at three different places on the surface spaced of few mm. After the monolayer deposition, we acquired once again three spectra at three different places of the surface and we used a 2-layer model (substrate/SAM) to fit the measured data and to determine the SAM thickness. We employed the previously measured optical properties of the substrate (background), and we fixed the refractive index of the organic monolayer at 1.50.[14] The three spectra measured on the sample were fitted separately using each of the three reference spectra, giving nine values for the SAM thickness. We calculated the mean value from these nine thickness values and the thickness incertitude corresponding to the standard deviation. Overall, we estimated the accuracy of the SAM thickness measurements at ± 2 Å.[13] Figure S5 and Table S1 summarize the measured values, compared to the geometry optimized length of the molecules (in gas phase, MOPAC, AM1 level). The measured thicknesses for the AH2 and AH3 SAMs are in good agreement with the simulated molecule length including the protecting group on the top of the SAM that remains since no deprotecting agent was used for the SAM formation. This suggests that the SAMs are closely packed. In the case of the AH1 SAM, the measured thickness is smaller than the molecule length, which suggests that the AH1 molecules in the SAM are slightly less compactly organized.

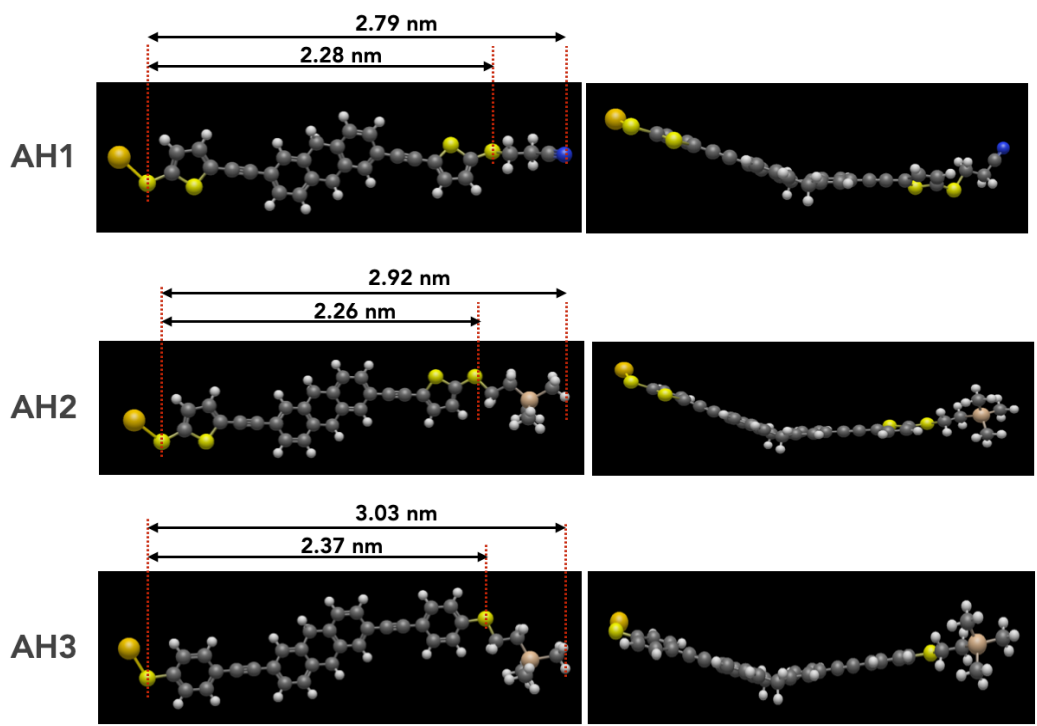

*Figure S5.* Optimized geometry of the three molecules (in gas phase, MOPAC, AM1 level), top and side views.

|  | ellipsometry (nm) | theoretical length (nm) -S to S- | theoretical length (nm) -S to end group |
|---|---|---|---|
| AH1 | 2.3 ± 0.2 | 2.28 | 2.79 |
| AH2 | 3.1 ± 0.2 | 2.26 | 2.92 |
| AH3 | 2.8 ± 0.2 | 2.37 | 3.03 |

*Table S1*. Measured thickness of the SAMs and comparison with the molecule length. The S-to-S length is taken from the geometry optimization in gas phase. We also give the length assuming that the end groups remain on the top of the SAM surface after the chemical grafting since no deprotection agent was used.

### 2.4. X-ray photoemission spectroscopy (XPS) and ultraviolet photoemission spectroscopy (UPS).

High resolution XPS spectra (Physical Electronics 5600 spectrometer fitted in an UHV chamber with a residual pressure of $3 \times 10^{-10}$ mbar) were recorded with a monochromatic Al$_{K\alpha}$ X-ray source (h$\nu$ = 1486.6 eV), a detection angle of 45° as referenced to the sample surface, an analyzer entrance slit width of 400 μm and with an analyzer pass energy of 12 eV. Background was subtracted by the Shirley method.[15] The peaks were decomposed using Voigt functions and a least squares minimization procedure. Binding energies were referenced to the C 1s BE, set at 284.8 eV. The same equipment was used for UPS characterizations with an ultraviolet source (He1, h$\nu$ = 21.2 eV).

The XPS spectra of the AH1, AH2 and AH3 SAMs show the expected elements. In particular, the S 2p region shows the expected two doublets (S $2p_{1/2}$ and S $2p_{3/2}$) associated to the sulfur bonded to gold, S-Au (S $2p_{1/2}$ at 162.3 eV, S $2p_{3/2}$ at 163.5 eV) and the "unbonded" to gold S atoms, *i.e.*, S-C (S $2p_{1/2}$ at 163.6 eV, S $2p_{3/2}$ at 164.9 eV) (Fig. S6). These doublets are separated by ca. 1.2-1.3 eV as expected with an amplitude ratio [S $2p_{1/2}$]/[S $2p_{3/2}$] set at 1/2. The same results were obtained for the three AH1, AH2 and AH3 samples (Fig. S6, values in Table S2). The S-Au bonding is confirmed by the calculated amplitude ratios (integrated peak area) of the non-bonded S atoms and the bonded S to Au [S-C]/[S-Au]. We obtain [S-C]/[S-Au] = 3.1 (for AH1 and AH2), and 1.15 (for AH3), in good agreement with the theoretical ratio of 3 (AH1 and AH2) and 1 for AH3. These results indicate that chemisorbed SAMs are formed via S-Au bonds whatever the chemical nature of the protecting groups and the nature of the π-conjugated side groups. Albeit it is likely that a certain amount of protecting groups remain at the upmost surface of the SAMs, the N 1s (SAM AH1) and Si 2p (SAMS AH2 and AH3) peaks were not detected due to the sensitivity limit.

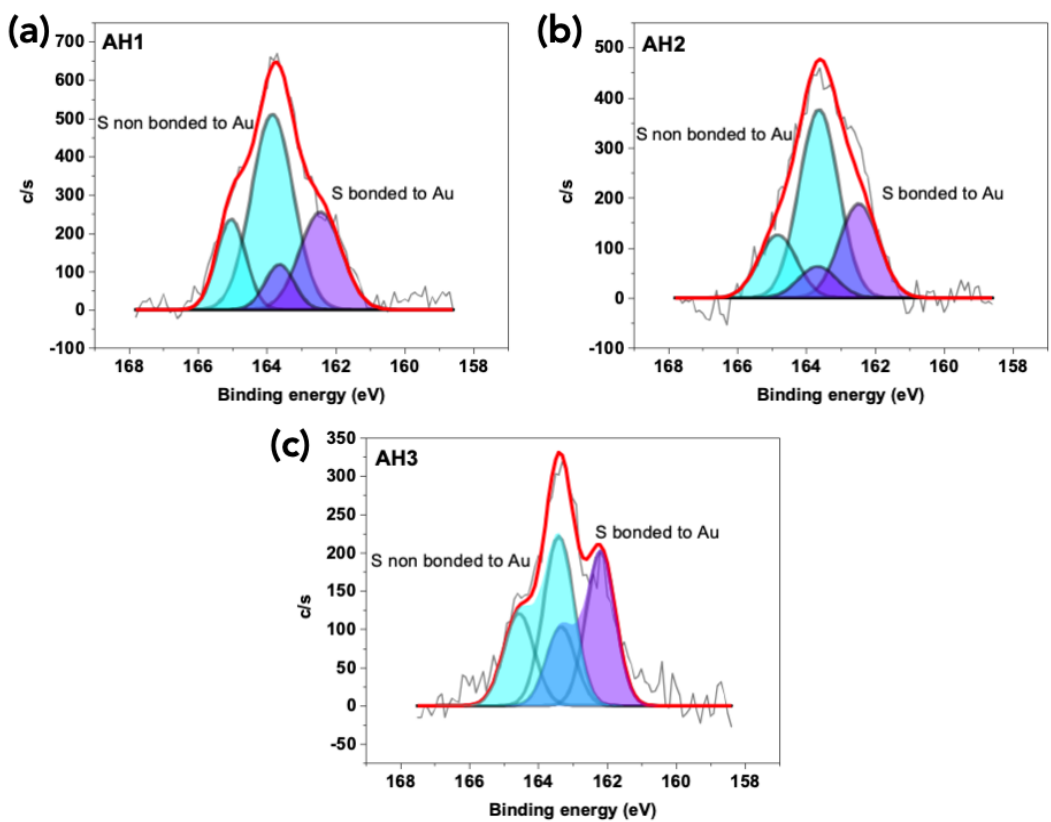

*Figure S6.* XPS spectra, S 2p region, of (a) the AH1 SAM, (b) the AH2 SAM and (c) the AH3 SAM. The dark lines are the measured data, the red lines are the fits with the deconvolution of the two doublets: the S bonded to Au (purple) and the S non bonded to Au (light blue).

|  | AH1 | AH2 | AH3 |
|---|---|---|---|
| S bonded to Au |  |  |  |
| S $2p_{1/2}$ (eV) | 162.3 | 162.3 | 162.2 |
| S $2p_{3/2}$ (eV) | 163.5 | 163.6 | 163.4 |
| S non bonded to Au |  |  |  |
| S $2p_{1/2}$ (eV) | 163.6 | 163.6 | 163.5 |
| S $2p_{3/2}$ (eV) | 164.9 | 164.8 | 164.7 |
| [S-C]/[S-Au], meas. | 3.1 | 3.1 | 1.15 |
| [S-C]/[S-Au], theory | 3 | 3 | 1 |

*Table S2.* Binding energies of the S 2p peaks. Peak area ratios (integrated peak area) of the non-bonded S atoms and the bonded S to Au [S-C]/[S-Au].

The UPS spectra (Fig. S7a) show that the onset of the HOMO is similar for the AH1 and AH2 SAMs, as expected, and located at about 0.89 eV below Au Fermi level. For the AH3 SAM, the HOMO is closer to the Fermi energy at 0.33 eV. This shift of the HOMO of the AH3 SAM is qualitatively in agreement (same trend) with the energy shift deduced from the analysis of the I-V curves (energy $\varepsilon_0$ is 0.35 to 0.47 eV for AH1 and AH2 MJs, Figs. 4, S11 and S12, and 0.14 eV for AH3 MJ, Fig. S13). The slight difference observed for the energy values of the HOMO from the I-V measurements is not seen from the UPS experiment, which is likely less sensitive. A more quantitative comparison is delicate since in the UPS measurements, only one electrode (Au substrate) is present, and in the I-V measurements, the second top electrode (C-AFM tip), can also induce change of the HOMO position due to an additional interface dipole and charge transfer. The secondary cutoff (Fig. S7b) shows the work functions of the naked $^{TS}$Au electrodes and the three $^{TS}$Au/SAM samples. The SAMs induce a reduction of the work function as usually observed due to the molecule and molecule/metal interface dipoles.[16-18] The WFs of the $^{TS}$Au/SAMs is around 4.2-4.4 eV (Fig. S7b). These values again confirm the grafting of the SAMs. A more detailed analysis is delicate and would require to know the molecular dipoles and also the precise density and organization of the molecules in the SAMs.[16]

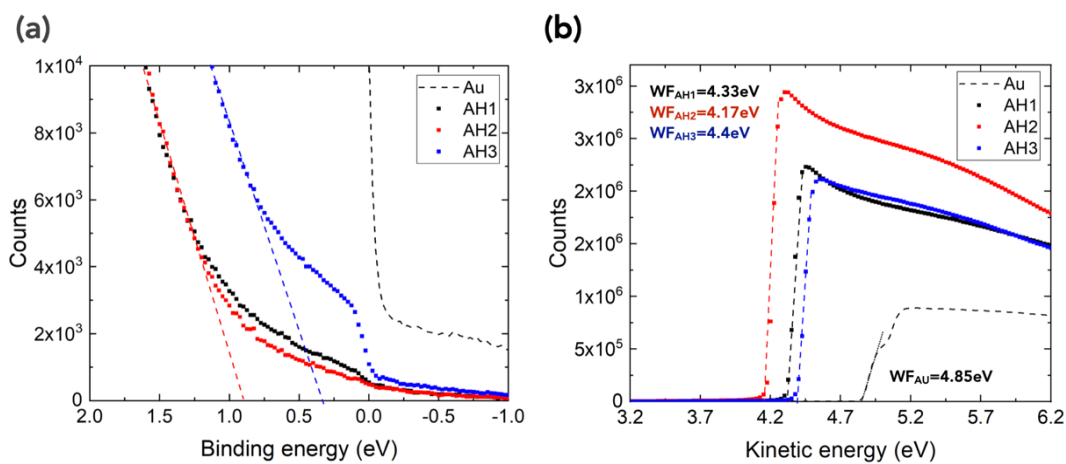

***Figure S7.*** *(a) UPS spectra of the AH1, AH2 and AH3 SAMs at the onset of the HOMO. The Au Fermi energy is set at 0 eV (black dashed line) from an UPS measurement on a clean reference Au substrate. The spectra are almost identical for the AH1 and AH2 SAMs, the red dashed line shows the extrapolation to zero used to determine the onset of the HOMO at 0.89 eV below the Au Fermi energy. The blue dashed line is the same extrapolation to determine the HOMO of the AH3 SAM at 0.33 eV. We note that in that case, the underlying Au electrode is more visible in the UPS spectrum. (b) Secondary cutoff to determine the work functions.*

## 3. Conductive AFM protocols and methods.

### 3.1. Conductive AFM measurements and THz irradiations.

The current−voltage characteristics were measured by conductive atomic force microscopy (Icon, Bruker), using PtIr coated tip (RMN-12PT400B from Bruker, 0.3 N/m spring constant). The 30 THz QCL laser was placed near the Icon and the THz beam was focused on the sample below the tip using an optical fiber (focus spot diameter = 300 µm). The emission wavelength of the QCL was set at 966.5 cm$^{-1}$ or 29.3THz (continuous wave), and the output power was adjustable. To form the molecular junction, the conductive tip was located at a stationary contact point on the SAM surface at a controlled loading force (6-8 nN). The voltage was applied on the substrate, the tip is grounded via the input of the current-voltage preamplifier. The C-AFM tip is located at different places on the sample (typically on an array - 10x10 grid - of stationary contact points spaced of 50-100 nm), at a fixed loading force and the I–V characteristics were acquired directly by varying the voltage for each contact point. The I-V characteristics were not averaged between successive measurements and typically hundreds I-V measurements were acquired on each sample. The load force was set at ≈6-8 nN for all the I-V measurements, a lower value leading to too many contact instabilities during the I-V measurements. The tip contact area was estimated (6-7.5 nm$^2$) with a mechanical model (*vide infra*, section 3.3).

For the 2.5 THz laser irradiation, we used a Bruker Multimode C-AFM mounted on the optical bench. The THz source is a $CO_2$ pumped methanol gas laser emitting at 2.522 THz (continuous wave) and its power was adjustable. To adapt the beam size to the dimension of the internal optics of the bench, the beam size is reduced with a set of two confocal parabolic mirrors (focus spot diameter = 160 μm). Then, the beam enters the optical bench and it is focused on the tip-sample junction with another parabolic mirror. In addition, an infrared laser was aligned with the THz laser, and a camera was used to observe the sample, giving a convenient view point to focalize the THz laser spot on the junction under the C-AFM tip. The same protocol as above was used to acquire the I-V datasets.

### *3.2. Data analysis*

Before to fit the I-V curves with the analytical models, the raw set of I-V data is analyzed and some I-V curves were discarded from the analysis:
- At high current, the I-V traces that reached the saturating current during the voltage scan (the compliance level of the trans-impedance amplifier).
- The traces displaying abrupt steps during the voltage scan (due to contact instabilities).

- At low current, the I-V traces that reached the sensitivity limit (almost flat I-V traces) and displayed noisy and random staircase behavior (due to the sensitivity limit of both the trans-impedance amplifier and the resolution of the ADC (analog-digital converter). The measurement yield for the three samples is summarized in Table S3.

|         | Total I-V acquired | I-Vs showing NDC | percentage |
|---------|--------------------|------------------|------------|
| AH1     | 400                | 116              | 29         |
| AH1(#2) | 92                 | 39               | 42         |
| AH2     | 56                 | 18               | 32         |
| AH3     | 30                 | 15               | 50         |

***Table S3***. *Statistical data on the total number of measured I-Vs and the percentage showing a clear NDC effect.*

For the histograms (Figs. 4, S11 – S14) of the fitted parameters with Eqs. (S4) and (S6), we have fitted all the I-Vs shown in the datasets (Figs. 2, S8 – S10), we have inspected all the fits and discarded those not fitting the data (bad coefficient of determination $R^2<0.9$ and fitted I-Vs not matching the data I-Vs).

### *3.3. Loading force and C-AFM tip contact area.*

The load force was set at ≈ 6-8 nN for all the I-V measurements, a lower value leading to too many contact instabilities during the I-V measurements. Considering: (i) the area per molecule on the surface (as estimated from the calculated geometry optimization), and (ii) the estimated C-AFM tip contact surface (see below), we estimate N, the number of molecules contacted under the C-AFM tip, as follows. As usually reported in literature[19-22] the contact radius, a, between the C-AFM tip and the SAM surface, and the SAM elastic deformation, δ, are estimated from a Hertzian model:[23]

$$a^2 = \left(\frac{3RF}{4E^*}\right)^{2/3} \tag{S1}$$

$$\delta = \left(\frac{9}{16R}\right)^{1/3}\left(\frac{F}{E^*}\right)^{2/3} \tag{S2}$$

with F the tip load force (6-8 nN), R the tip radius (20 nm) and E* the reduced effective Young modulus defined as:

$$E^* = \left(\frac{1}{E^*_{SAM}} + \frac{1}{E^*_{tip}}\right)^{-1} = \left(\frac{1-\nu^2_{SAM}}{E_{SAM}} + \frac{1-\nu^2_{tip}}{E_{tip}}\right)^{-1} \tag{S3}$$

In this equation, $E_{SAM/tip}$ and $\nu_{SAM/tip}$ are the Young modulus and the Poisson ratio of the SAM and C-AFM tip, respectively. For the Pt/Ir (90%/10%) tip, we have $E_{tip}$ = 204 GPa and $\nu_{tip}$ = 0.37 using a rule of mixture with the known material data.[24] These parameters for the present SAM are not known and, in general, they are not easily determined in such a monolayer material. Thus, we consider the value of an effective Young modulus of the SAM $E^*_{SAM}$ ≈ 38 GPa as determined for the "model system" alkylthiol SAMs from a combined mechanic and electron transport study.[21] With these parameters, we estimate a ≈ 1.4-1.5 nm (contact area ≈ 6-7.5 nm$^2$) and δ ≈ 0.09-0.12 nm. With a molecular packing density of about 1 molecule per 30Å$^2$ (estimated from the theoretical conformation optimization, the molecule fits inside a cylinder with a diameter of ∼ 6 Å), we infer that about 20-25 molecules are measured in the MJs.

### 3.4. Fits of the I-V curves with analytical models.

The I-V curves with a NDC behavior were fitted by a two-site model. The current-voltage behavior is given by:[25, 26]

$$I(V) = N\frac{e}{h}\frac{\Gamma(2\tau)^2}{\Delta^2 + \Gamma^2}\left[arctan\left(\frac{\frac{1}{2}eV_i - \varepsilon_1}{\Gamma/2}\right) + arctan\left(\frac{\frac{1}{2}eV_i + \varepsilon_1}{\Gamma/2}\right) + \frac{\Gamma}{2\Delta}ln\left(\frac{\left(\frac{1}{2}eV_i - \varepsilon_1\right)^2 + (\Gamma/2)^2}{\left(\frac{1}{2}eV_i + \varepsilon_1\right)^2 + (\Gamma/2)^2}\right)\right.$$

$$\left. +arctan\left(\frac{\frac{1}{2}eV_i - \varepsilon_2}{\Gamma/2}\right) + arctan\left(\frac{\frac{1}{2}eV_i + \varepsilon_2}{\Gamma/2}\right) - \frac{\Gamma}{2\Delta}ln\left(\frac{\left(\frac{1}{2}eV_i - \varepsilon_2\right)^2 + (\Gamma/2)^2}{\left(\frac{1}{2}eV_i + \varepsilon_2\right)^2 + (\Gamma/2)^2}\right)\right]$$

(S4)

with e the electron charge, h the Planck constant, Γ the coupling energy between the electrodes and the molecule, τ the coupling energy between the left and right π-moieties of the molecules, and N set to 25 the number of molecules contacted by the C-AFM tip (*vide supra*). For simplicity, we used the same Γ parameter on the two sides albeit the electrodes are not similar (Au and PtIr). This is not critical since all the I-Vs are almost symmetric with the voltage polarity. The parameters $\varepsilon_1$ and $\varepsilon_2$ are voltage-dependent effective energy positions of the molecular orbitals defined by $\varepsilon_1 = \varepsilon_0 - \Delta/2$, $\varepsilon_2 = \varepsilon_0 + \Delta/2$, with $\varepsilon_0$ the molecular orbital energy of the π-moieties with respect to the Fermi energy of the electrodes and Δ is the energy shift between the two orbitals under the application of a voltage V given by:[25]

$$\Delta = \sqrt{(\alpha eV)^2 + 4\tau^2}$$

(S5)

with α the fraction of the applied voltage that drops inside the molecule. This parameter is evaluated by two approaches. (i) the theoretical MO splitting (Fig. S16) is fitted with Eq. (S5), given α=0.79 and τ=10.1 meV; (ii) considering the energy position of the degenerated HOMO/HOMO-1 theoretically calculated at 0.47/0.49 eV (at zero bias), we fixed $\varepsilon_0$=0.48 eV in Eq. (S4) and fitted the I-Vs (dataset AH1, Fig. 2a) with α as an adjustable parameter (plus Γ and τ). We get α values between 0.72 and 0.88. Thus, we used α=0.8 to fit all the I-V datasets. This value was also confirmed by simulating the voltage drop inside the MJ (*vide infra*, section 5, Fig. S18). A similar value was reported for the MCBJ experiment.[25]

The I-V curves with no NDC were fitted with a single energy level (SEL) model. This model assumes that only one molecular orbital, at an energy $\varepsilon_{0-THz}$ from the Fermi energy of the electrodes, contributes to the electron transport.[27,28]

$$I(V) = N \frac{8e}{h} \frac{\Gamma_1 \Gamma_2}{\Gamma_1 + \Gamma_2} \left[ \arctan\left( \frac{\varepsilon_{0-THz} + \frac{\Gamma_1}{\Gamma_1+\Gamma_2} eV}{\Gamma_1 + \Gamma_2} \right) - \arctan\left( \frac{\varepsilon_{0-THz} - \frac{\Gamma_2}{\Gamma_1+\Gamma_2} eV}{\Gamma_1 + \Gamma_2} \right) \right] \quad (S6)$$

where $\Gamma_1$ and $\Gamma_2$ are the electronic coupling energies of the molecule to the two electrodes. Note that we have obtained $\Gamma_1 \approx \Gamma_2$ (Fig. S14), justifying the use of a single parameter $\Gamma$ in Eq. (S4) above. This model is valid at 0 K, since the Fermi-Dirac electron distribution of the electrodes is not taken into account. However, it was shown that it can be reasonably used (within an error of a few percent) to fit data measured at room temperature for voltages within a limited range. The condition of applicability of the 0K SEL model to room temperature experimental data is given by a numerical analysis reported in Ref. [29]. In our case, we verified *a posteriori*, that with $\varepsilon_0 \approx 0.5$-$0.6$ eV, $\Gamma_1$ and $\Gamma_2$ around 1 meV (Fig. S14), this condition is |V|< 0.94 V), and thus the fits done here (Fig. S14) between -1 and 1 V are reasonably accurate.

All the fits were done with the routine included in ORIGIN software, using the method of least squares and the Levenberg Marquardt iteration algorithm.

## 4. Additional data.

### *4.1. Terahertz switching*

To check the reproducibility of the THz switching behavior, we repeated the same measurements on another AH1 sample from another batch: AH1(#2), Fig. S8.

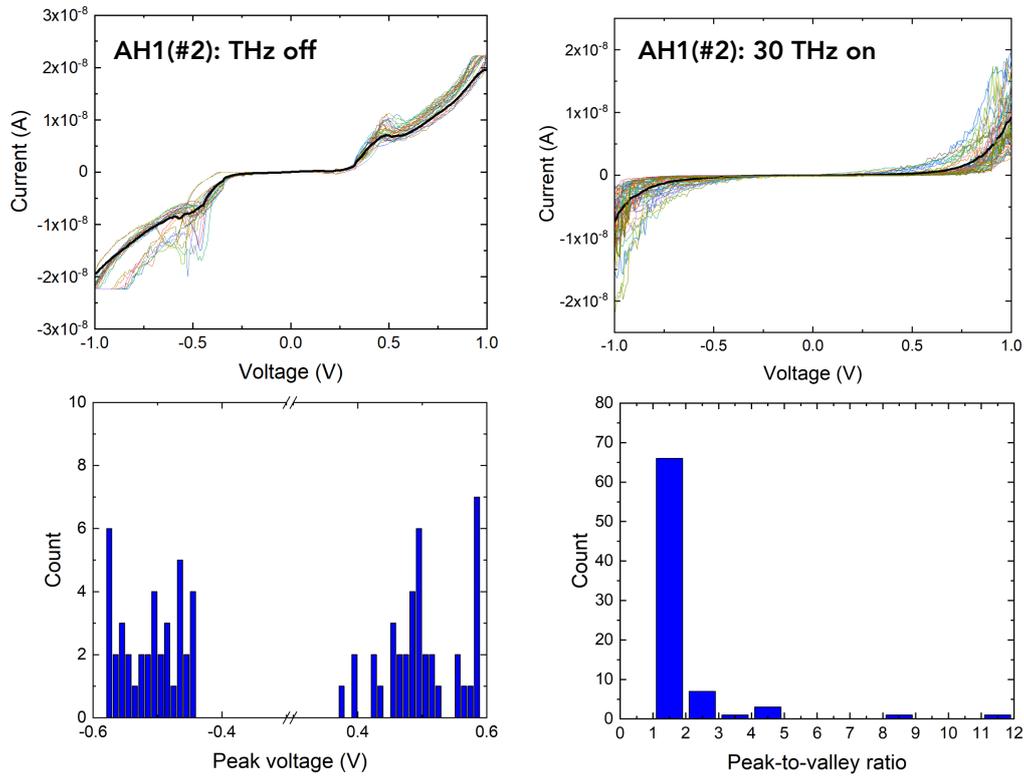

*Figure S8*. (a) Current-voltage dataset for the $^{TS}$Au-AH1(#2)//PtIr tip MJs without THz waves irradiation (39 I-V curves, different colors, acquired at different places on the SAM). The bold black line is the mean $\bar{I}$-V curve (b) Current-voltage (52 I-V curves, different colors) dataset for the $^{TS}$Au-AH1(#2)//PtIr tip MJs under a 30 THz continuous wave irradiation (at a power of 40 mW). The bold black line is the mean $\bar{I}$-V curve. (c) Histogram of the voltages of the peak currents (from the dataset in panel (a)) for the positive, $V_{p+}$, and negative, $V_{p-}$, voltages. (d) Histogram of the peak-to-valley ratios (from the dataset in the panel (a)) calculated as for every I-V.

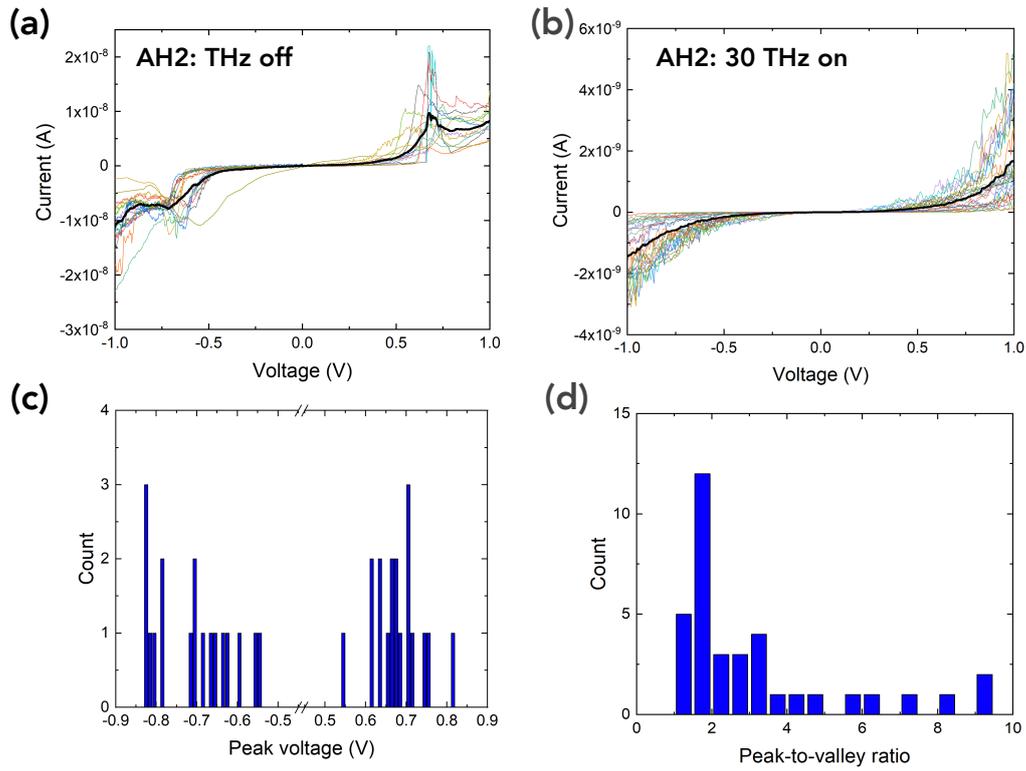

*Figure S9.* *(a) Current-voltage dataset for the $^{TS}$Au-AH2//PtIr tip MJs without THz waves irradiation (18 I-V curves, different colors, acquired at different places on the SAM). The bold black line is the mean Ī-V curve (b) Current-voltage (27 I-V curves, different colors) dataset for the $^{TS}$Au-AH2//PtIr tip MJs under a 30 THz continuous wave irradiation (at a power of 40 mW). The bold black line is the mean Ī-V curve. (c) Histogram of the voltages of the peak currents (from the dataset in panel (a)) for the positive, $V_{p+}$, and negative, $V_{p-}$, voltages. (d) Histogram of the peak-to-valley ratios (from the dataset in panel (a)) calculated for every I-V*

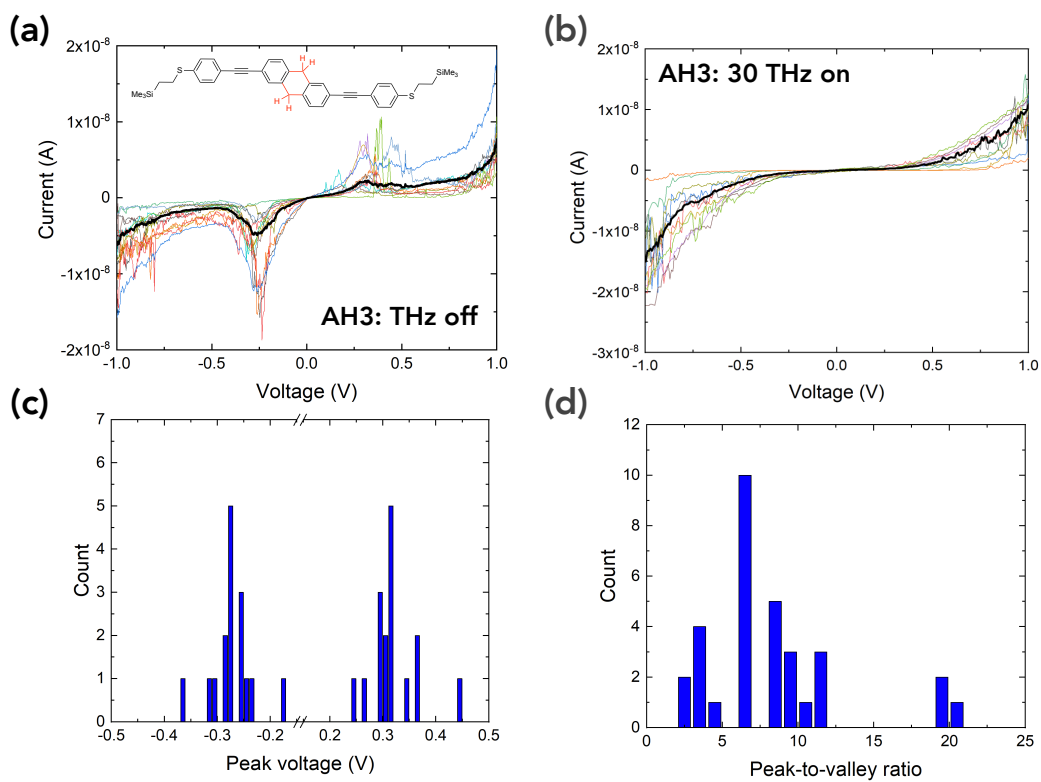

***Figure S10.*** *(a) Current-voltage dataset for the $^{TS}$Au-AH3//PtIr tip MJs without THz waves irradiation (15 I-V curves, different colors, acquired at different places on the SAM). The bold black line is the mean $\bar{I}$-V curve (b) Current-voltage (11 I-V curves, different colors) dataset for the $^{TS}$Au-AH3//PtIr tip MJs under a 30 THz continuous wave irradiation (at a power of 40 mW). The bold black line is the mean $\bar{I}$-V curve. (c) Histogram of the voltages of the peak currents (from the dataset in panel (a)) for the positive, $V_{p+}$, and negative, $V_{p-}$, voltages. (d) Histogram of the peak-to-valley ratios (from the dataset in panel (a)) calculated as for every I-V.*

We verified that the THz switching effect does not depend on the chemical nature of the left and right π-moieties. We also synthesized the 2,6-bis((4-((2-(trimethylsilyl)ethyl)thio)phenyl)ethynyl)-9,10-dihydroanthracene molecule (AH3, inset of Fig. S10) for which NDC behavior was already reported for a single molecule using mechanically controlled break junctions (MCBJ).[25] We clearly reproduced the NDC behavior with the <sup>TS</sup>Au-AH3//PtIr tip MJs (Fig. S10a). Two NDC peaks were observed at about +/- 0.3 V (Fig. S10c). This gap between the two peaks (~ 0.6 V) is consistent with the MCBJ results. This feature corresponds to a HOMO (at zero volt) located at ~ 0.14 eV with respect to the Fermi energy of the electrodes (Fig. S13) according to the fit with the two-site model, in qualitative agreement with the MCBJ results (at zero bias, degenerate HOMO and HOMO-1 close to the electrode Fermi energy).[25] The smaller value of $\varepsilon_0$ for the AH3 MJ is consistent with the UPS measurements (see Fig. S7). Under the irradiation with a 30 THz wave, the NDC behavior is suppressed as for the AH1 and AH2 molecular junctions (Fig. S10b).

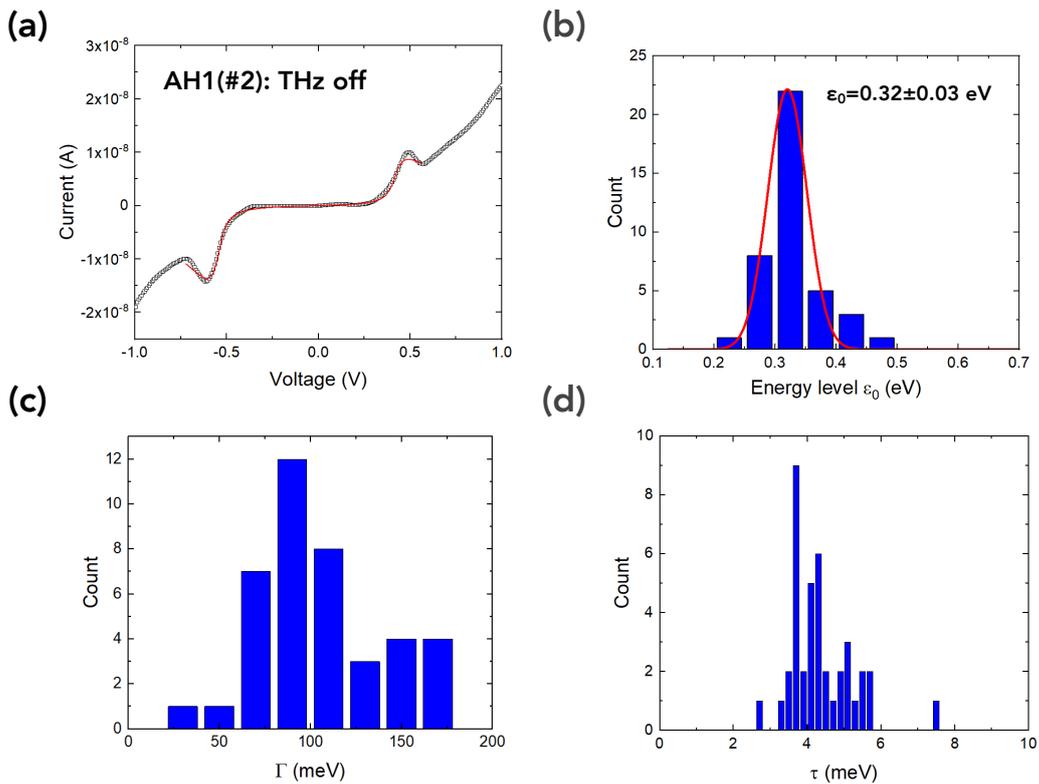

*Figure S11.* (a)Typical fit of the two-site model on a I-V trace of the <sup>TS</sup>Au-AH1(#2)//PtIr tip MJ (dataset of Fig. S8a). The fit parameters are: (a) $\varepsilon_0$=0.38 eV, Γ=139 meV, τ=5.2 meV. The fits are limited between the voltage positions of the two current valleys since the monotonous increase of the background "off-resonance" current with the voltage at higher voltages is not taken into account by this model (see Fig. 4d in Ref. [25]). (b-d) Histograms of the fit parameters obtained by fitting all individual I-V curves of the dataset (Fig. S8a): (b) energy of molecular orbital $\varepsilon_0$, the red line is the fit with a Gaussian distribution; (c) electrode coupling energy $\Gamma$ and (d) intramolecular electronic coupling energy $\tau$.

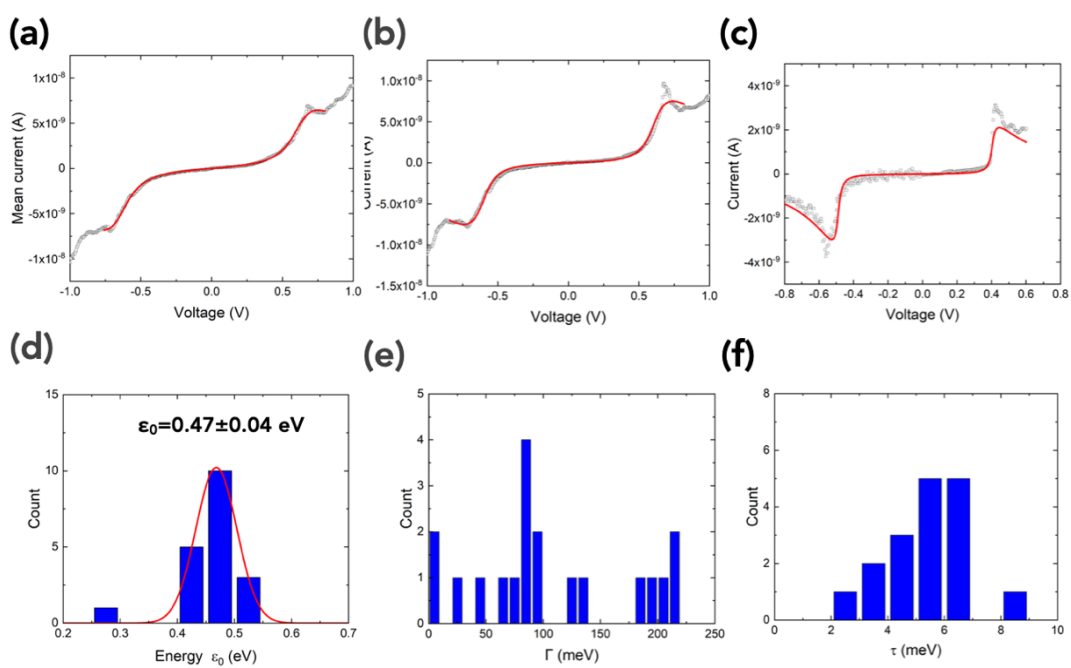

*Figure S12*. Fits of the two-site model on (a) the mean $\bar{I}$-V of the $^{TS}$Au-AH2//PtIr tip MJ (dataset of Fig. S9a) and (b-c) two typical I-V from the dataset with a marked NDC behavior. The fit parameters are: (a) $\varepsilon_0$=0.52 eV, $\Gamma$=205 meV, $\tau$=3.92 meV; (b) $\varepsilon_0$=0.44 eV, $\Gamma$=129 meV, $\tau$=4.8 meV; (c) $\varepsilon_0$=0.32 eV, $\Gamma$=79 meV, $\tau$=2.3 meV. The fits are limited between the voltage positions of the two current valleys since the monotonous increase of the background "off-resonance" current with the voltage at higher voltages is not taken into account by this model (see Fig. 4d in Ref. [25]). (d-f) Histograms of the fit parameters obtained by fitting all individual I-V curves of the dataset (Fig. S9a): (d) energy of molecular orbital $\varepsilon_0$, the red line is the fit

*with a gaussian distribution; (e) electrode coupling energy Γ and (f) intramolecular electronic coupling energy τ.*

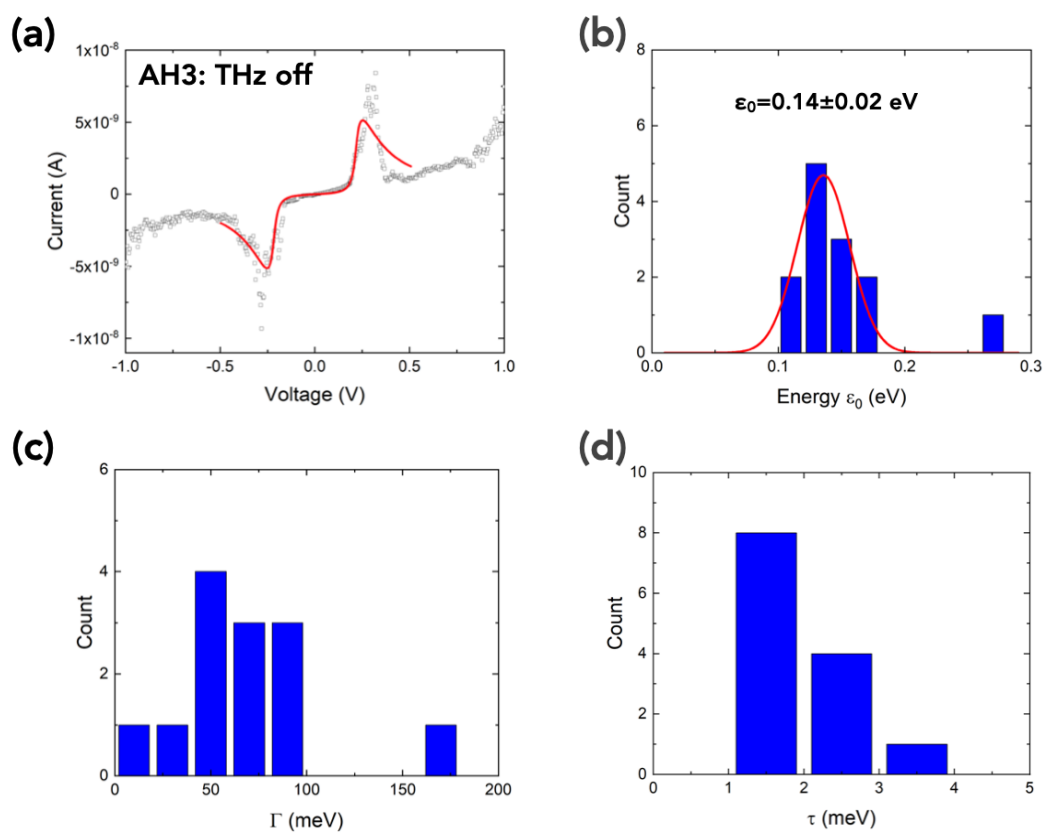

***Figure S13.*** *(a)Typical fit of the two-site model on a I-V trace of the $^{TS}$Au-AH3//PtIr tip MJ (dataset of Fig. S10a). The fit parameters are: (a) $ε_0$=0.16 eV, Γ=29 meV, τ=2.7 meV. The fits are limited between the voltage positions of the two current valleys since the monotonous increase of the background "off-resonance" current with the voltage at higher voltages is not taken into account by this model (see Fig. 4d in Ref. [25]). (b-d) Histograms of the fit parameters obtained by fitting all individual I-V curves of the dataset (Fig. S10a): (b) energy of molecular orbital $ε_0$, the red line is the fit with a Gaussian distribution; (c) electrode coupling energy Γ and (d) intramolecular electronic coupling energy τ.*

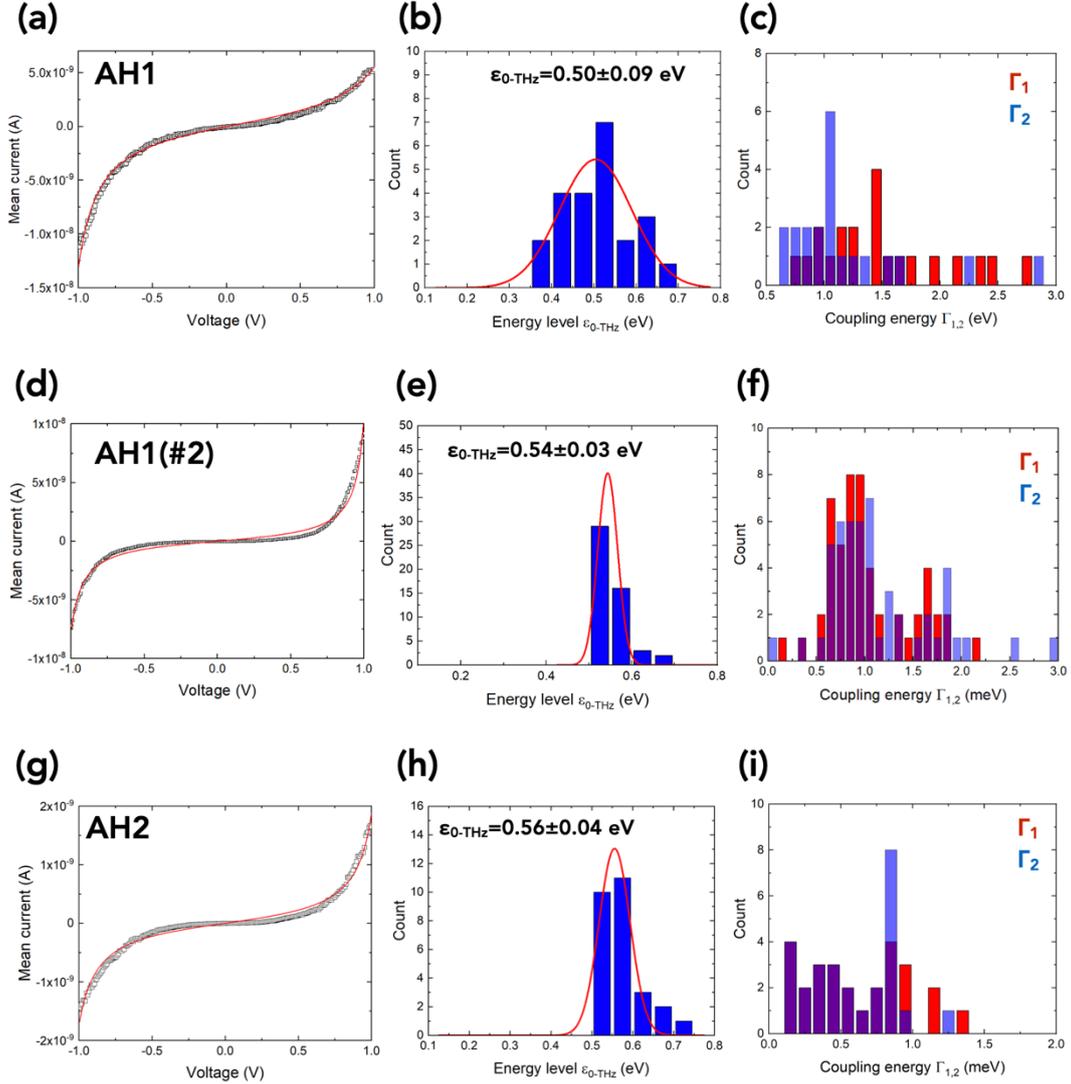

*Figure S14*. (a) Fit of the SEL model (Eq. S6, red line) on the mean Ī-V curve (AH1 junction, from Fig. 2b) with the parameters: $\varepsilon_{0-THz}$=0.47 eV, $\Gamma_1$= 2.0 meV and $\Gamma_2$=1.7 meV. (b) Histograms of the $\varepsilon_{0-THz}$ values from the fit of all individual I-V traces in the dataset of Fig. 2b. The histogram is fitted by a gaussian distribution with $\varepsilon_{0-THz}$=0.50±0.09 eV. (c) Histograms of the fitted electrode coupling energies. $\Gamma_1$ and $\Gamma_2$ span in the range ≈ 0.5 to 3 meV. (d) Fit of the SEL model (Eq. S6, red line) on the mean Ī-V curve (AH1(#2) junction, from Fig. S8b) with the parameters: $\varepsilon_{0-THz}$=0.53 eV, $\Gamma_1$= 0.9 meV and $\Gamma_2$=1.0 meV. (e) Histograms of the $\varepsilon_{0-THz}$ values from the fit of all individual I-V traces in the dataset of Fig. S8b. The histogram is fitted by a *gaussian distribution with $\varepsilon_{0\text{-}THz}$=0.54±0.03 eV. (f) Histograms of the fitted electrode coupling energies. $\Gamma_1$ and $\Gamma_2$ span in the range ≈ 0.5 to 3 meV. (g) Fit of the SEL model (Eq. S6, red line) on the mean $\bar{I}$-V curve (AH2 junction, from Fig. S9b) with the parameters: $\varepsilon_{0\text{-}THz}$=0.56 eV, $\Gamma_1$= 0.59 meV and $\Gamma_2$=0.61 meV. (h) Histograms of the $\varepsilon_{0\text{-}THz}$ values from the fit of all individual I-V traces in the dataset of Fig. S9b. The histogram is fitted by a gaussian distribution with $\varepsilon_{0\text{-}THz}$=0.56±0.04 eV. (i) Histograms of the fitted electrode coupling energies. $\Gamma_1$ and $\Gamma_2$ span in the range ≈ 0.1 to 1.5 meV.*

### 4.2. Plasmonic effect

As a control experiment, we measured I-Vs under 30 THz irradiation (same power of 40 mW) of MJs made of SAMs of oligo(phenylene ethynylene) (OPE). We did not observe any modification of the I-Vs (Fig. S15), especially no current increase due to PAT (plasmon assisted tunneling) like for the background currents of the AH1, AH2 and AH3 MJs (Figs. 2, S8 to S10). We conclude that the field confinement in the tip/SAM/surface plasmonic cavity is not strong enough to generate a significant PAT current in our experimental. For example, the generation of PAT current was observed at 2.5 THz and 4K in a C60 molecular junction with bowtie-shaped antenna electrodes to enhance the coupling efficiency between the THz wave and the tunneling electrons in the MJs.[30] In our case, this experimental observation would require more elaborate calculations of the charge transfer plasmon resonance in the THz range as done in reference [31] to explain the experimental observation of tunneling charge transfer plasmon in ethanedithiolate and 1,4-benzenedithiolate MJs.[32]

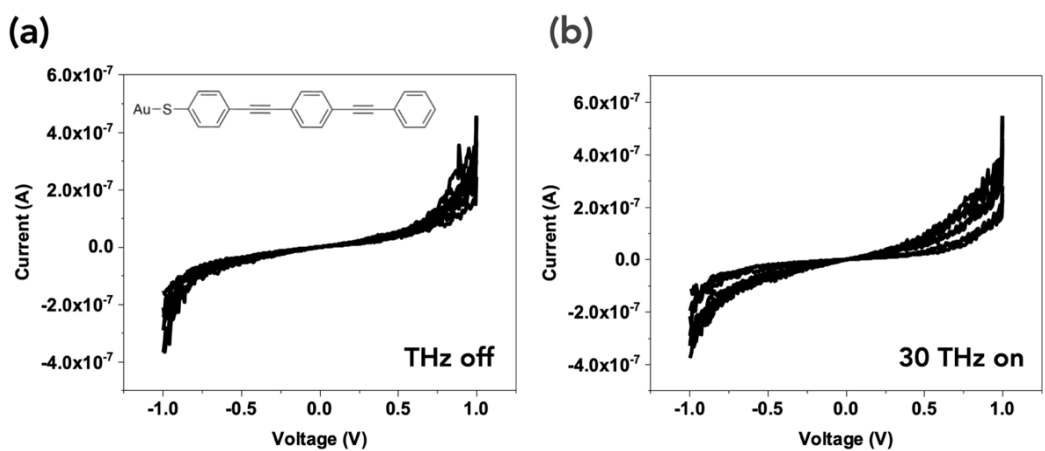

*Figure S15. Current-voltage curves (10 traces) measured for a $^{TS}$Au-OPE3//PtIr tip: (a) without THz irradiations, (b) under 30 THz irradiation at P=40 mW.*

## 5. Theory

### 5.1. DFT and NEGF calculations.

The geometric structure of the molecule was first optimized in the gas phase at the DFT level, with the B3LYP functional[33] and a 6-31 G (d,p) basis set[34] with the Gaussian09 software.[35] The relaxed molecule is then sandwiched between two (111) gold electrode through a sulfur anchoring atom. The geometry of the interfaces was then optimized by relaxing the molecule and the top two gold layers until forces are below 0.02 eV/Å. A density mesh cutoff of 100 Ha and a (3×3×1) Monkhorst Pack k- sampling were used for the relaxation. Once the geometry of the molecular junction is optimized, a layer of gold ghost atoms has been added on the top layer of the gold electrodes at a distance of 1.7 Å away so that the work function of the clean Au (111) surface of 5.25 eV matches the experimental value[36] and previous theoretical studies.[16, 37] We note that albeit a PtIr tip is used for the measurements, a gold planar electrode is used for the calculations as validated in a previous work.[38] The calculations of the electronic properties were performed using the QuantumATK/2022.03 package[39, 40] with the GGA.PBE exchange-correlation functional and a SZP (DZP) basis set for valence gold electrons (other valence electrons).[41] The core electrons were frozen and described by the norm-conserving Troullier–Martins pseudopotentials.[42] The electronic transmission spectra of the molecular junctions are calculated at different biases and the I-V curves are simulated using the Landauer-Büttiker formalism[43] within the coherent transport regime (more details in the Supplementary Information). For transport calculations, we used a (5*5*100) Monkhorst Pack k- sampling, a mesh cutoff of 100 Hartree and a temperature of 300 K. These parameters have been carefully tested to ensure the convergence of the transmission spectrum.

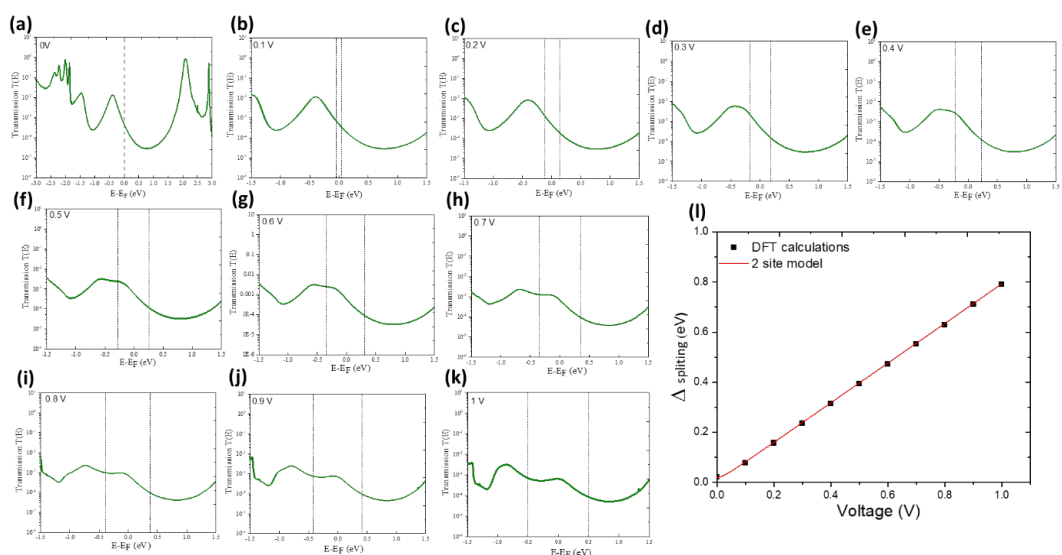

*Figure S16. (a-k) Details of the calculated transmission probability T(E) for the Au-AH1-AU MJ from 0 to 1V, step 0.1V. (l) Energy difference Δ between the HOMO and HOMO-1 versus the applied voltage (dark squares) and fit (red line) by Eq. S5 with α = 0.79 and τ = 10.1 meV.*

### 5.2. Au-AH1//Au molecular junction

For the MJ with the molecule chemisorbed at only one interface, the same approach was used, except that for the second gold electrode added on the top side of the molecular layer (Fig. S17a), a van der Waals contact is assumed between the molecular layer and the top electrode. We used an interatomic distance determined as the sum of the van der Waals radii of the sulfur and gold atoms (3.46 Å).

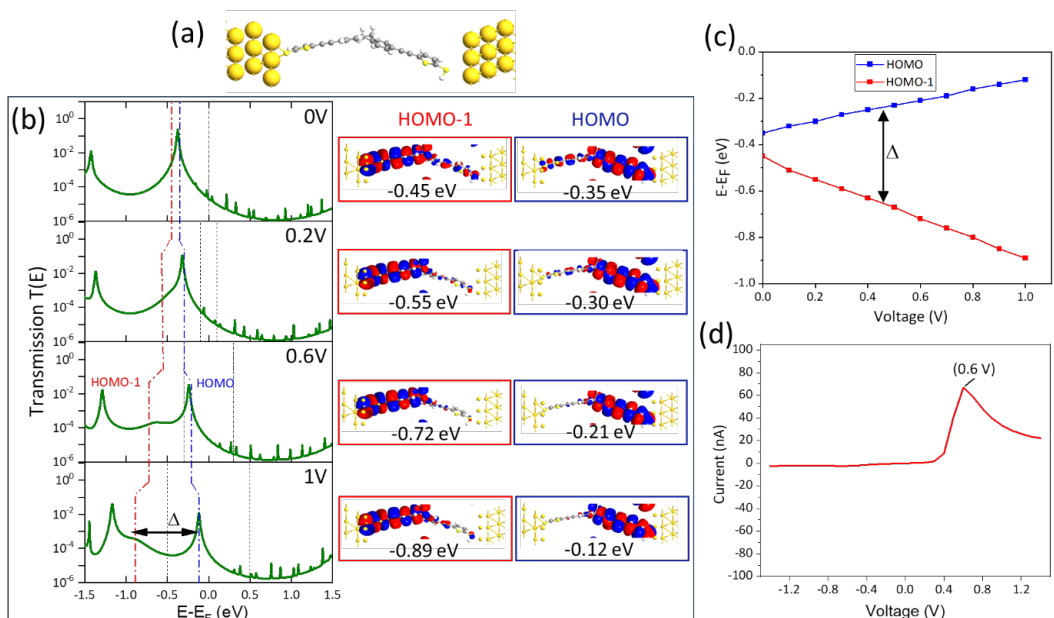

*Figure S17.* *(a) Optimized geometry of the Au-AH1//Au molecular junctions. (b) Calculated transmission coefficient, T(E), at several applied voltages. The blue and red dashed lines highlight the voltage dependent shift of the energy position of the HOMO and HOMO-1, respectively. The vertical dark dashed line indicated the position of the Fermi energy of the two electrodes (energy window used to calculate the current). The blue and red boxes show the corresponding localization of the HOMO and HOMO-1 orbitals, with their energy position with respect to the Fermi energy of the electrode. (c) Evolution of the energy gap Δ between the HOMO and HOMO-1 with the voltage (V>0). At V<0, no NDC peak is calculated because both the HOMO and HOMO-1 are pulled out from the energy window. (d) Simulated I-V curves with only a NDC peak at +0.6 V. We note that the peak-to-valley ratio increases to ≈ 3.5 (vs. ≈ 1.3. in the case of the chemisorbed Au-AH1-Au*

*MJ at the two sides, Fig. 5, main text). This is due to a decrease in the off-resonant current ("background" I-V) induced by the weak coupling at the mechanical contact.*

### 5.3. Junctions under bias

In order to simulate the molecular junctions under bias, the transmission spectra have been first calculated out of equilibrium for each bias in a self-consistent way, that is, the electronic structure of the entire junction is re-optimized for every applied bias. Then, the I-V characteristics have been calculated based on the Landauer-Büttiker formalism,[43] that links the transmission spectrum to the current in a coherent transport regime by using the I-V curve function implemented in QuantumATK/2022.03.[39, 40] In fact, when a bias is applied, the current is calculated via the integration of the transmission spectrum within a bias window defined by a Fermi-Dirac statistics in the left and right electrodes:

$$I(V) = \frac{2e}{h} \int T(E) \left[ f\left(\frac{E - \mu_R}{k_B T_R}\right) - f\left(\frac{E - \mu_L}{k_B T_L}\right) \right] dE \quad \text{(S7)}$$

Where T(E) is the transmission spectrum, calculated at each bias, E the incident electron energy, f the Fermi function, $\mu_{R/L} = E_F \pm (eV/2)$, the chemical potential of the right/left electrode, with $E_F$ the metal Fermi energy, e the elementary charge, and V the applied bias. $T_{R/L}$ the temperature of the right/left electrode set here to 300K, $k_B$ the Boltzmann constant and h the Planck constant.

### 5.4. Voltage drop

The voltage drop is the difference in real space in effective potential $\delta V_E(r,V)$ between two self-consistent calculations, *i.e.* $\delta V_E(r,V_1) - \delta V_E(r,V_0)$ where the reference calculation "$V_0$" is associated to zero bias. The obtained quantity shows how the applied bias "$V_1$" is distributed across the device, *i.e.* where the voltage drops. In this work, we have calculated the voltage drop for the Au-AH1-Au and Au-AH1//Au junctions under their peak voltage; 0.7V and 0.6V, respectively. The results reveal that most of the voltage drop takes place around the molecule, not near the electrodes (Fig. S18).

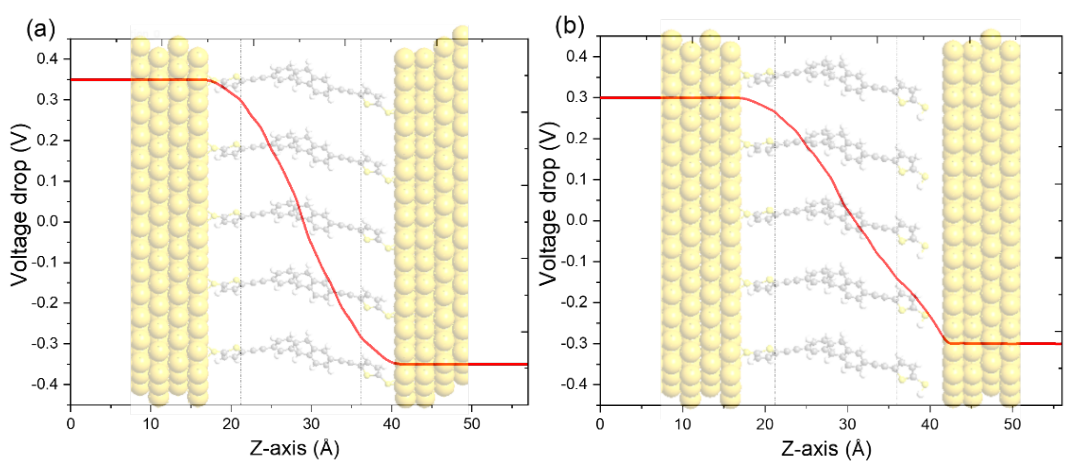

*Figure S18. The calculated total voltage drop across the device projected on molecular axis (Z axis) in (a) the Au-AH1-Au junction under 0.7V and in (b) the Au-AH1//Au junction under 0.6V (i.e. at peak voltage). A voltage of 0.58 V (0.41 V) is dropped inside the molecule part marked by the dashed lines for the Au-AH1-Au (Au-AH1//Au) junction, which corresponds to α≈0.83-0.68 in good agreement with the determination by fitting the molecular orbital splitting, Δ, with Eq. (S5) (see Fig. S16).*

### 5.5. Lorentzian fitting: Γ broadening of HOMO and HOMO-1 transmission peaks

To estimate the coupling between the AH1 molecule and the electrodes (Γ), we fitted the width of the HOMO and HOMO-1 transmission peak with a Lorentzian function.

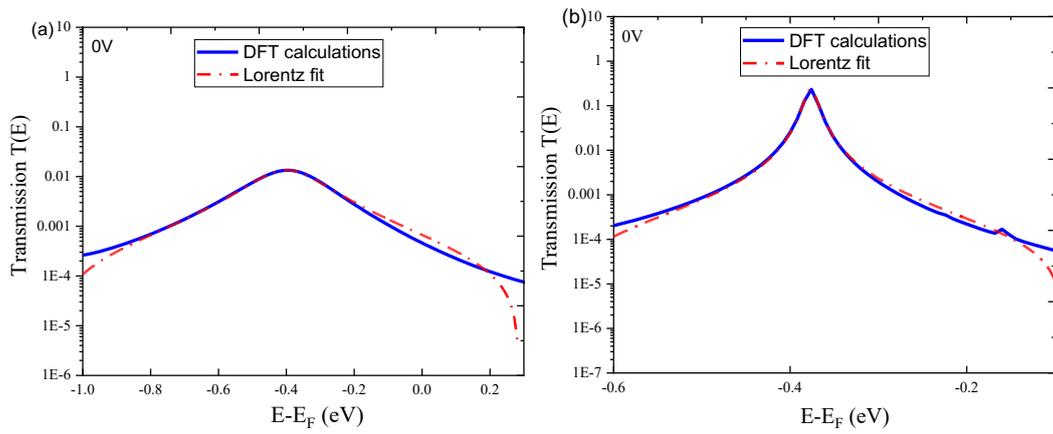

***Figure S19.*** *Lorentzian fitting of the transmission peak at zero bias of (a) Au-AH1-Au junction and (b) Au-AH1//Au junction. The fitted Γ indicates that the Au-AH1-Au exhibits larger broadening (227 meV) compared to Au-AH1//Au (159 meV).*

*5.6. Calculated transmission spectra T(E) at V<0 for the Au-AH1//Au junction*

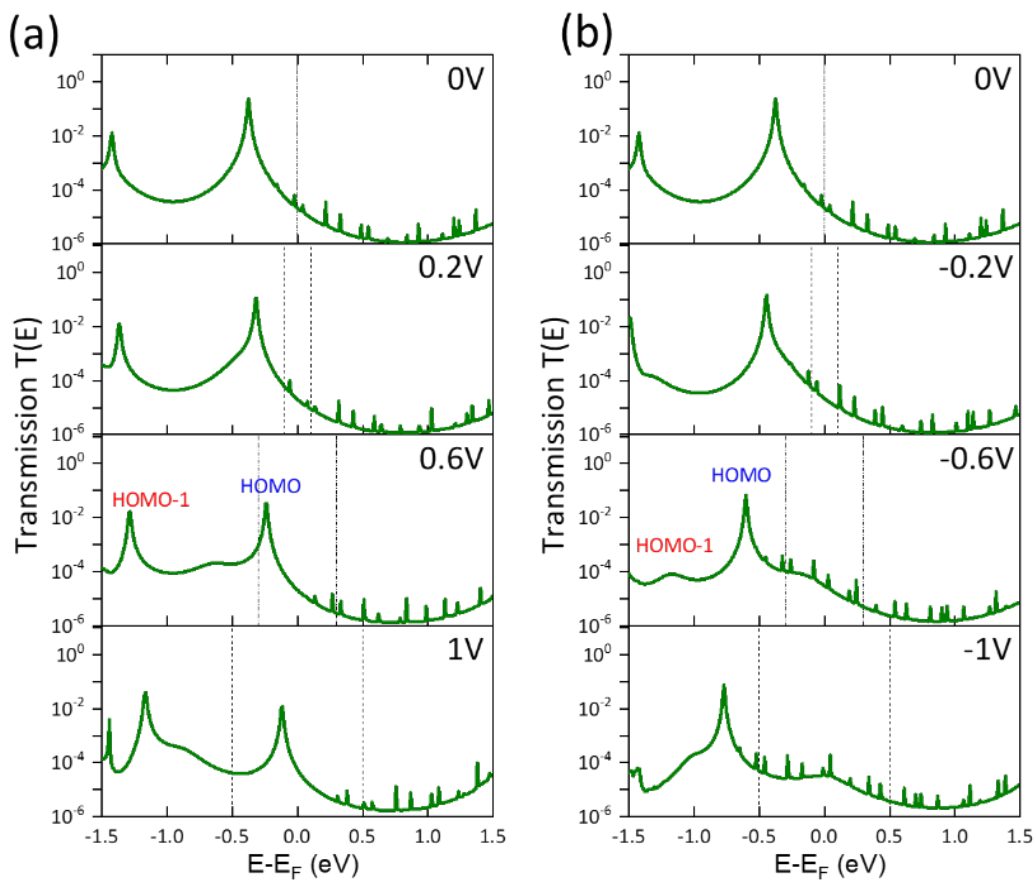

***Figure S20***. *Calculated transmission spectra of the Au-AH1//Au junction for (a) positive and (b) negative voltages. Due to the asymmetry of the junction, the transmission spectra at V<0 shift toward the chemisorbed electrode (left electrode) and thus the HOMO and HOMO-1 peaks are pulled out from the energy window for negative voltages giving rise to a negligible current for all voltages without the possibility to catch the NDC behavior at V<0.*